\documentclass[a4paper,11pt]{article}
\pdfoutput=1 
\usepackage{jheppub}
\usepackage{amsmath, amsthm, amssymb, bbm, dsfont,relsize}
\usepackage{slashed}

\def\Bbb{\mathbb}

\font\teneurm=eurm10 \font\seveneurm=eurm7  \font\fiveeurm=eurm5
\newfam\eurmfam
\textfont\eurmfam=\teneurm \scriptfont\eurmfam=\seveneurm
\scriptscriptfont\eurmfam=\fiveeurm

\font\teneusm=eusm10 \font\seveneusm=eusm7 \font\fiveeusm=eusm5
\newfam\eusmfam
\textfont\eusmfam=\teneusm \scriptfont\eusmfam=\seveneusm
\scriptscriptfont\eusmfam=\fiveeusm

\font\tencmmib=cmmib10 \skewchar\tencmmib='177
\font\sevencmmib=cmmib7 \skewchar\sevencmmib='177
\font\fivecmmib=cmmib5 \skewchar\fivecmmib='177
\newfam\cmmibfam
\textfont\cmmibfam=\tencmmib \scriptfont\cmmibfam=\sevencmmib
\scriptscriptfont\cmmibfam=\fivecmmib

\def\16{{\bf 16}}
\def\1{{\bf 1}}
\def\2{{\bf 2}}
\def\3{{\bf 3}}
\def\4{{\bf 4}}

\def\t{\widetilde}
\def\tilde{\widetilde}
\def\R{{\Bbb{R}}}
\def\hat{\widehat}

\font\teneurm=eurm10 \font\seveneurm=eurm7 \font\fiveeurm=eurm5
\newfam\eurmfam
\textfont\eurmfam=\teneurm \scriptfont\eurmfam=\seveneurm
\scriptscriptfont\eurmfam=\fiveeurm

\font\teneusm=eusm10 \font\seveneusm=eusm7 \font\fiveeusm=eusm5
\newfam\eusmfam
\textfont\eusmfam=\teneusm \scriptfont\eusmfam=\seveneusm
\scriptscriptfont\eusmfam=\fiveeusm

\font\tencmmib=cmmib10 \skewchar\tencmmib='177
\font\sevencmmib=cmmib7 \skewchar\sevencmmib='177
\font\fivecmmib=cmmib5 \skewchar\fivecmmib='177
\newfam\cmmibfam
\textfont\cmmibfam=\tencmmib \scriptfont\cmmibfam=\sevencmmib
\scriptscriptfont\cmmibfam=\fivecmmib

\numberwithin{equation}{section}
\def\d{\mathrm d}
\def\bar{\overline}
\def\C{{\Bbb C}}

\def \Q {{\mathcal Q}}
\def \S {{\mathcal S}}

\def \N {{\mathcal N}}

\def \F {{\mathcal F}}
\def \hF {{\hat{\mathcal F}}}
\def \S {{\mathcal S}}
\def \R {{\mathcal R}}
\def \T {{\mathcal T}}
\def \W {{\mathcal W}}
\def \H {{\mathcal H}}
\def \F {{\mathcal F}}
\def \hF {{\hat{\mathcal F}}}
\def \J {{\mathcal J}}
\def \Y {{\mathcal Y}}
\def \O {{\mathcal O}}
\def \vPh {{\mathbf\Phi}}
\def \I {{\mathcal I}}
\def \P {{\mathcal P}}

\title{\boldmath Chiral algebras in Landau-Ginzburg models}

\author[1]{Mykola Dedushenko} 

\affiliation[1]{Joseph Henry Laboratories, Princeton University, Princeton NJ USA 08540}

\emailAdd{dedushenko@gmail.com}

\abstract{Chiral algebras in the cohomology of the $\bar{Q}_+$ supercharge of two-dimensional $\N=(0,2)$ theories on flat spacetime are discussed. Using the supercurrent multiplet, we show that the answer is renormalization group invariant for theories with an R-symmetry. For $\N=(0,2)$ Landau-Ginzburg models, the chiral algebra is determined by the operator equations of motion, which preserve their classical form, and quantum renormalization of composite operators. We study these theories and then specialize to the $\N=(2,2)$ models and consider some examples.}

\begin{document}
\maketitle
\flushbottom

\section{Introduction}
Two-dimensional theories with $\N=(0,2)$ supersymmetry have been attracting attention over the last couple of decades. A motivation largely came from their potential phenomenological relevance for heterotic string compactifications, which require the internal theory to be an $\N=(0,2)$ SCFT. But these theories are interesting and rich quantum field theories by themselves, which makes them a good object to study and apply various physical ideas. Thinking in that direction, gauge theories are of course of particular importance in theoretical physics and deserve attention in various dimensions and with various amounts of supersymmetry. But besides that, $\N=(0,2)$ gauged linear sigma models are known to be a useful tool to construct $\N=(0,2)$ SCTFs, and hence heterotic string vacua, as infrared (IR) fixed points of the renormalization group (RG) flow (see \cite{WPh, DK} or just \cite{Dis} and references therein).

Recently, the dynamics of two-dimensional $\N=(0,2)$ supersymmetric gauge theories, both abelian and non-abelian, have seen an increasing interest, especially due to developments in \cite{GGP1, GGP2, GGP3}. At the same time, more basic models of $\N=(0,2)$ interacting matter without gauge fields, sometimes referred to as $\N=(0,2)$ Landau-Ginzburg (LG) models, have been studied, some references being \cite{DK, Mel1, Mel2}. One can think of starting from an $\N=(0,2)$ LG model with global flavor symmetries and then gauging these global symmetries to obtain an $\N=(0,2)$ gauge theory. Therefore, it might be beneficial for certain questions to first study properties of LG models and then ask what happens to these properties after gauging.

The property we want to study in this paper is the chiral algebra in the cohomology of one of the supercharges. The supercharges $Q_+$ and $\bar{Q}_+$ of the $(0,2)$ theory satisfy: 
\begin{equation}
\{Q_+, \bar{Q}_+\}=2P_{++},
\end{equation}
where $P_{++}$ is the right-moving translation generator. Since $\bar{Q}_+^2=0$, one can study its cohomology, and the above equation implies that it is holomorphic, or, in terms of light-cone coordinates $x^{\pm\pm}$ (in Lorenzian signature), the cohomology depends non-trivially only on $x^{--}$, while differentiation with respect to $x^{++}$ annihilates cohomology classes. This observation was first made in \cite{VaV} and then in \cite{W} used to elucidate some properties of $\N=(2,2)$ LG models and their IR fixed points. Then, part of the analysis from \cite{W} was extended to $\N=(0,2)$ gauge theories in \cite{SW}. 

Chiral algebras of $\N=(0,2)$ half-twisted sigma models were studied to some extent in the literature due to their connection with the theory of chiral differential operators. In particular, the perturbative approach was developed in \cite{W02} and \cite{Tan}, and some non-perturbative aspects were studied in \cite{TanY} and \cite{Y}. There was also a number of papers on topological rings (which are finite sectors of chiral algebras in $\mathcal{N}=(0,2)$ theories), some examples being \cite{Katz:2004nn, Guffin:2008kt, Guffin:2008pi, Guo:2015gha}. However, it seems that systematic analysis of chiral algebras in $\N=(0,2)$ LG models and gauge theories has not been performed yet. Our goal is to make a small step in this direction.

In this paper we first study some general properties of $\N=(0,2)$ theories on $\Bbb{R}^{1,1}$. Then we restrict to a certain class of models, namely LG models, and later consider LG models with $\N=(2,2)$ supersymmetry and give some examples.

An important question about chiral algebras of $\N=(0,2)$ theories is how they behave under the RG flow. It turns out that in theories with the R-symmetry chiral algebra is an RG invariant, while it is not completely clear whether this is the case for more general theories without R-symmetry. The RG invariance underlies all approaches to extract some useful information about the CFT in the IR, such as, for example, in \cite{W} and \cite{SW}. It seems that this has never been proved in the literature though. 

We study RG invariance of the chiral algebra on general grounds using the $\N=(0,2)$ supercurrent multiplet described in \cite{DS}. If the theory has an R-symmetry, the supercurrent multiplet becomes what is called the R-multiplet. In such a situation, only using manipulations with the R-multiplet, we show that there is a stress-energy tensor in the cohomology. This fact underlies the finding of \cite{SW} that the stress-energy tensor in the cohomology is not spoiled by anomaly if and only if the R-symmetry is non-anomalous. So we obtain conformal symmetry in the chiral algebra. It is interesting to note that this stress-energy tensor is the one of the half-twisted $\N=(0,2)$ theory. Since the stress-energy tensor in cohomology is identified with the left-moving stress-energy tensor of the CFT in the IR, we can say the following: in $\N=(0,2)$ theories with the R-symmetry, the RG flow from the UV to the IR performs a half-twist. 

Having conformal symmetry in the cohomology is a very strong restriction. It turns out that because of it, the chiral algebra cannot depend on any dimensionful constants. This fact allows one to argue that for the LG models, the chiral algebra is in fact tree level exact, and the OPE of the cohomology classes can be computed using the free field correlators. This makes chiral algebra a potentially powerful tool for obtaining exact results. We should note, however, that in this discussion we assume that there are no non-perturbative effects. This seems a reasonable assumption for LG models on a topologically trivial space with a topologically trivial target only.

In Section 2 of this paper we discuss general aspects of $\N=(0,2)$ theories, namely the supercurrent multiplet and its ambiguities, conformal invariance in the cohomology and its implications for the OPE. We also discuss what is the chiral algebra of SCFT's and relate it to the notion of a chiral ring of \cite{LVW}. We then discuss the superspace technique to describe the $\bar{Q}_+$ cohomology. 

In Section 3 we discuss general properties of $\N=(0,2)$ LG models and specify to the quasi-homogeneous superpotentials. We then review the statement that the OPE of the cohomology classes can be computed using the free theory and argue that the chiral algebra is tree level exact (but still different from the classical algebra due to the singularities one encounters in defining composite operators). In Section 4 we specify to $\N=(2,2)$ supersymmetry, and in Section 5 we discuss a few examples, focusing on the $A_{k+1}$ series of $\N=2$ minimal models. We then conclude in Section 6 and mention some further directions.

\section{$\N=(0,2)$ theories}
In this section we discuss some general aspects of two-dimensional $(0,2)$-supersymmetric theories and their chiral algebras. 
\subsection{Conventions and some generalities}
The two-dimensional theories with $(0,2)$ supersymmetry are characterized by the existence of two conserved supercharges $Q_+$ and $\bar{Q}_+$ of positive (or right-handed) chirality acting on the Hilbert space of the theory. They satisfy:
\begin{align}
Q_+^2 = \bar{Q}_+^2&=0,\cr
\{Q_+,\bar{Q}_+\}&= 2P_{++},
\end{align}
where $2P_{++}=P_0 + P_1$ is a light-cone momentum. The standard geometric realization of supersymmetry is to consider the superspace $\Bbb R^{2|2}$ with bosonic coordinates $x^0, x^1$ and fermionic coordinates $\theta^+$ and $\bar\theta^+$. Superfields are distributions on this superspace taking values in operators acting on the Hilbert space. The supercharges $Q_+$ and $\bar{Q}_+$ act on operators (and therefore on superfields) by commutators, and the geometric realization of this action is through the differential operators:
\begin{align}
\Q_+ &= \frac{\partial}{\partial\theta^+} + i\bar\theta^+\frac{\partial}{\partial x^{++}},\cr
\bar\Q_+ &= -\frac{\partial}{\partial\bar\theta^+} - i\theta^+\frac{\partial}{\partial x^{++}},
\end{align}
so that for an arbitrary superfield $F$, we have $[Q_+,F]_\pm=\Q_+ F$, where $[\dots]_\pm$ denotes a graded commutator. These operators obviously satisfy the required relation $[\Q_+,\bar\Q_+]=-2i\frac{\partial}{\partial x^+}$. We also have another pair of differential operators on $\Bbb{R}^{2|2}$, $D_+$ and $\bar{D}_+$, given by:
\begin{align}
D_+ &= \frac{\partial}{\partial\theta^+} - i\bar\theta^+\frac{\partial}{\partial x^{++}},\cr
\bar{D}_+ &= -\frac{\partial}{\partial\bar\theta^+} + i\theta^+\frac{\partial}{\partial x^{++}},
\end{align}
for which the key property is that they anticommute with $\Q_+$ and $\bar\Q_+$ and hence can be used in constructing supersymmetric Lagrangians.

We also adopt the convention in which hermitian conjugation reverses the order of fermions, that is $(\theta_1\theta_2)^\dagger=\bar\theta_2 \bar\theta_1$.

The basic superfields are

\begin{itemize}
\item[1)] Chiral superfields satisfying $\bar{D}_+\Phi=0$. The component expansion contains a complex scalar $\phi$ and a left spinor $\psi_+$:
\begin{equation}
\Phi = \phi + i\theta^+\psi_+ - i\theta^+\bar\theta^+ \partial_{++}\phi
\end{equation}
The antichiral superfield satisfies $D_+\bar\Phi=0$ and is given by:
\begin{equation}
\bar\Phi=\bar\phi + i\bar\theta^+\bar\psi_+ + i\theta^+\bar\theta^+\partial_{++}\bar\phi.
\end{equation}

\item[2)] Fermi superfields satisfying $\bar{D}_+\Lambda=E(\Phi)$, where $E(\Phi)$ is a chiral superfield constructed as a holomorphic function of basic chiral superfields. The component expansion contains a right-handed spinor $\lambda$ and an auxiliary field $G$:
\begin{equation}
\Lambda = \lambda + \theta^+ G - i\theta^+\bar\theta^+\partial_{++}\lambda -\bar\theta^+ E(\Phi),
\end{equation}
where $E$ itself has to be expanded in components. The opposite chirality Fermi superfield satisfies $D_+\bar\Lambda=-\bar{E}(\bar\Phi)$ and is given by:
\begin{equation}
\bar\Lambda=\bar\lambda +\bar\theta^+\bar{G} +i\theta^+\bar\theta^+\partial_{++}\bar\lambda -\theta^+\bar{E}(\bar\Phi).
\end{equation}

\item[3)] Real superfield.
\end{itemize}

If $U$ is a real superfield, it can always be thought of as a real part of some chiral superfield (not necessarily a local one; also we will allow for superfields which are chiral only on-shell). We will denote the imaginary part of this chiral superfield by $\tilde{U}$. Then $U+i\tilde{U}$ is chiral on-shell and $U-i\tilde{U}$ is antichiral. The relation between $U$ and $\tilde{U}$ is:
\begin{align}
\label{mir}
\bar{D}_+ \tilde{U} &= i\bar{D}_+ U,\cr
D_+\tilde{U} &= - i D_+ U,
\end{align}
up to equations of motion. This $\tilde{U}$ is defined up to a term which is constant on-shell. If the component expansion of $U$ is
\begin{equation}
U = u + i\theta^+\chi_+ + i\bar\theta^+ \bar\chi_+ + \theta^+\bar\theta^+\partial_{++}v,
\end{equation}
where we wrote the highest component as a derivative of some function $v$, then the component expansion of $\tilde{U}$ is:
\begin{equation}
\tilde{U}= v +\theta^+\chi_+ - \bar\theta^+\bar\chi_+ - \theta^+\bar\theta^+\partial_{++}u,
\end{equation}
again up to terms which vanish on equations of motion.

Note that if we want components of $U$ and $\tilde{U}$ to be local operators, then $U$ cannot be an arbitrary local real superfield. Its highest component, written as $\partial_{++}v$ above, should be a derivative of a local field. Only in such a case $v$ above is also local and hence $\tilde{U}$ is also the local superfield.

\subsection{Supercurrent multiplet and RG invariance}\label{sec_supcurr}
\subsubsection{General case}
The general $\N=(0,2)$ multiplet containing the stress-energy tensor and the supersymmetry current was described in \cite{DS}. It is referred to as the supercurrent multiplet. It consists of real superfields $\S_{++}, \T_{----}$ and a complex superfield $\W_-$ satisfying\footnote{Our conventions are different from \cite{DS}}:
\begin{align}
\label{constr}
\partial_{--}\S_{++} &= D_+\mathcal{W}_- - \bar{D}_+\bar{\mathcal W}_-,\cr
\bar{D}_+\T_{----} &= \partial_{--}\mathcal{W}_-,\cr
D_+\T_{----} &= \partial_{--}\bar{\mathcal W}_-,\cr
\bar{D}_+\W_- &= C,
\end{align}
where $C$ is a complex constant (a space-filling brane current). The component expansions which solve these constraints are:
\begin{align}
\label{components}
\S_{++}&=j_{++} - 2i\theta^+S_{+++} - 2i\bar\theta^+\bar{S}_{+++} -2\theta^+\bar\theta^+ T_{++++},\cr
\W_- &= -\bar{S}_{+--} - i\theta^+ \left(T_{++--} + \frac{i}{2}\partial_{--}j_{++} \right) - \bar\theta^+ C + i\theta^+\bar\theta^+\partial_{++}\bar{S}_{+--},\cr
\bar\W_- &= -S_{+--} + i\bar\theta^+ \left(T_{++--} - \frac{i}{2}\partial_{--}j_{++} \right) - \theta^+ \bar{C} - i\theta^+\bar\theta^+\partial_{++}S_{+--},\cr
\T_{----} &= T_{----} - \theta^+\partial_{--}S_{+--} + \bar\theta^+\partial_{--}\bar{S}_{+--} + \frac{1}{2}\theta^+\bar\theta^+ \partial^2_{--}j_{++}.
\end{align}
Applying constraints \eqref{constr} to these expansions implies conservation of $S_+$ (the supersymmetry current), conservation of $T$ (the stress-energy tensor) and symmetry of $T$:
\begin{align}
\partial_{++}S_{+--} + \partial_{--}S_{+++} &=0,\cr
\partial_{++}T_{\pm\pm --} + \partial_{--} T_{\pm\pm ++}&=0,\cr
T_{++--} - T_{--++}&=0.
\end{align}
Quite naturally, constraints \eqref{constr} do not determine the supercurrent multiplet uniquely. There are two types of ambiguities which preserve both the conservation laws and the form of equations \eqref{constr}. One ambiguity corresponds to improvement transformations:
\begin{align}
\label{improve}
\S_{++} &\to \S_{++} + [D_+,\bar{D}_+]U,\cr
\W_- &\to \W_- + \partial_{--}\bar{D}_+ U,\cr
\bar\W_- &\to \bar\W_- + \partial_{--}D_+ U,\cr
\T_{----} &\to \T_{----} + \partial_{--}^2 U,
\end{align}
where $U$ is an arbitrary real scalar superfield. These transformations do not change conserved charges.

Another ambiguity corresponds to the possibility of modifying the supercurrent multiplet by another conserved current (say, corresponding to some flavor symmetry), satisfying an additional requirement of locality which will be explained in a moment. If we have another conserved superspace current $\I_{\pm\pm}$, that is a pair of real superfields satisfying:
\begin{equation}
\partial_{--}\I_{++} + \partial_{++}\I_{--}=0,
\end{equation}
then we can use it to shift the supercurrent multiplet, i.e. define a new multiplet:
\begin{align}
\S_{++} &\to \tilde{\S}_{++}=\S_{++} + \I_{++},\cr
\W_- &\to \tilde\W_{-}=\W_- + \frac{i}{2}\bar{D}_+\I_{--},\cr
\bar\W_- &\to \tilde{\bar\W}_{-}=\bar\W_- - \frac{i}{2}D_+\I_{--},\cr
\T_{----} &\to \tilde\T_{----}=\T_{----} + \frac{1}{2}\partial_{--}\tilde\I_{--}.
\end{align}
Note that in the last equation we use $\tilde\I_{--}$, a real superfield related to $\I_{--}$ as in \eqref{mir}. That is, $\tilde{\I}_{--}$ is such that $\I_{--}+i\tilde{\I}_{--}$ is chiral. The new superfields $\tilde\S_{++}$, $\tilde\W_-$ and $\tilde\T_{----}$ will also satisfy the constraints \eqref{constr}. However, most conserved charges will be shifted by this transformation. Note that for the above transformation to make sense in a local QFT, both $\I_{\pm\pm}$ and $\tilde\I_{--}$ have to be local, so there is an extra requirement on $\I_{\pm\pm}$ that not only it has to be a conserved local superspace current, but also $\tilde\I_{--}$ has to be local. In the cases of interest for us, this will actually be the case.

One can easily read off the action of $\bar{Q}_+$ on various components of the supercurrent multiplet, and we are interested in the following:
\begin{align}
\{\bar{Q}_+, S_{+++}\} &= -i \left(T_{++++} + \frac{i}{2}\partial_{++} j_{++} \right),\cr
\{\bar{Q}_+, S_{+--}\} &= i \left(T_{++--} - \frac{i}{2}\partial_{--} j_{++} \right),\cr
\{\bar{Q}_+, T_{++++}\} &= \partial_{++} \bar{S}_{+++},\cr
\{\bar{Q}_+, T_{++--}\} &= -\partial_{++}\bar{S}_{+--},\cr
\{\bar{Q}_+, T_{----}\} &= -\partial_{--}\bar{S}_{+--}.
\end{align}
We see that neither component of the stress-energy tensor is annihilated by $\bar{Q}_+$, so components of $T$ by itself do not represent any $\bar{Q}_+$-cohomology classes. However, certain relations hold in the cohomology, in particular $T_{++--} - \frac{i}{2}\partial_{--}j_{++}$ is $\bar{Q}_+$-exact.
If we define the ``virial current'' $V_\mu$ as:
\begin{equation}
V_{--}=0,\quad V_{++}=i j_{++},
\end{equation}
then we have:
\begin{equation}
T_\mu^\mu = \partial^\mu V_\mu - \{\bar{Q}_+,4iS_{+--}\},
\end{equation}
which looks like condition for an effective scale-invariance \cite{Polch}, with the effective current for constant dilatations given by $d_\mu = x^\nu T_{\nu\mu} - V_\mu$. This current is ``almost conserved'': 
\begin{equation}
\partial^\mu d_\mu = \{\bar{Q}_+,\dots\}.
\end{equation}
The current $d_\mu$ itself is not $\bar{Q}_+$-closed. Even though $d_\mu$ is not precisely conserved, only up to $\bar{Q}_+$-exact terms, we still can try to define a ``charge'' $D$ corresponding to this current. If we have a local operator $\mathcal{O}(0)$ inserted at the origin, we define the action of $D$ on this operator as follows. Pick a contour $C$ enclosing $\mathcal{O}(0)$ and define:
\begin{equation}
[D,\mathcal{O}(0)]=\oint_C \star d(x)\mathcal{O}(0) = \oint_C \d x^\mu \epsilon_{\mu\nu}d^\nu(x)\mathcal{O}(0).
\end{equation}
This definition is clearly contour-dependent, since $d_\mu(x)$ is not conserved. As we deform the contour a bit, $[D,\mathcal{O}(0)]$ changes by $[\partial^\mu d_\mu(x),\mathcal{O}]$ integrated over the area swept by the deformation of the contour. But $\partial^\mu d_\mu(x)$ is $\bar{Q}_+$-exact, so if $\mathcal{O}(0)$ is $\bar{Q}_+$-closed, the change in $[D,\mathcal{O}(0)]$ under the contour deformation is $\bar{Q}_+$-exact. This means that $[D,\mathcal{O}(0)]$ is well-defined up to a $\bar{Q}_+$-exact piece when it acts on $\bar{Q}_+$-closed operators. Moreover, one can check that:
\begin{equation}
[D,\bar{Q}_+]=\bar{Q}_+,
\end{equation}
which shows that $D$ maps $\bar{Q}_+$-closed operators into $\bar{Q}_+$-closed operators. So we conclude that $D$ is a well-defined operator in the cohomology. It generates scale-transformations there. Since $D$ is not $\bar{Q}_+$-closed itself, we can say that scale transformations act as outer automorphisms in the cohomology.

\subsubsection{Emergent conformal invariance in the cohomology}
In the previous subsection we considered a general $\N=(0,2)$ theory in 2d, which a priori did not have any R-symmetries. The lowest component $j_{++}$ of the superfield $\S_{++}$ did not satisfy any conservation laws and, moreover, was not even accompanied by $j_{--}$. As was noted in \cite{DS}, if we restrict to the case $C=0$ and $\W_-=\frac{i}{2}\bar{D}_+\R_{--}$, where $\R_{--}$ is another real superfield (and also relabel $\S_{++}$ by $\R_{++}$), we get what is called an R-multiplet. The equation relating $\S_{++}$ and $\W_-$ becomes simply $\partial_{--}\R_{++} + \partial_{++}\R_{--}=0$, so the lowest component $j_{--}$ of $\R_{--}$ together with $j_{++}$ form a conserved R-current. So we have:
\begin{equation}
\R_{--}=j_{--} - 2i\theta^+ S_{+--} - 2i\bar\theta^+\bar{S}_{+--} - 2\theta^+\bar\theta^+ T_{++--},
\end{equation}
with $\partial_{++}j_{--} + \partial_{--}j_{++}=0$. In this situation, it becomes possible to define a new stress-energy tensor:
\begin{align}
\label{tw}
\tilde{T}_{++++}&=T_{++++} + \frac{i}{2}\partial_{++} j_{++},\cr
\tilde{T}_{++--}&=T_{++--} - \frac{i}{2}\partial_{--} j_{++},\cr
\tilde{T}_{----}&= T_{----} - \frac{i}{2}\partial_{--} j_{--},
\end{align}
which is also symmetric and conserved (by virtue of the conservation of $j$), but also it satisfies:
\begin{align}
\tilde{T}_{++++} &= \{\bar{Q}_+,\dots\},\quad \tilde{T}_{++--}=\{\bar{Q}_+,\dots\},\cr 
\tilde{T}_{----} &\ne \{\bar{Q}_+,\dots\},\quad \{\bar{Q}_+, \tilde{T}_{----}\}=0.
\end{align}
This procedure for $\N=(0,2)$ theories is known as a half-twisting. The above relations demonstrate that when it can be performed, one has the full 2d conformal invariance in the cohomology of $\bar{Q}_+$: the cohomology class represented by $\tilde{T}_{----}$ plays the role of the holomorphic\footnote{To be more precise, we should Wick rotate to the Euclidean signature in order to have holomorphy.} stress-energy tensor. It also ensures that the $\bar{Q}_+$-cohomology is invariant under the RG flow. The RG invariance of the chiral algebra implies that it carries a useful information about the IR fixed point.

Let us also take a closer look at the ambiguities of the supercurrent multiplet in the presence of R-symmetry. The improvement transformations are determined by a real superfield $U$ though \eqref{improve}, which tells us how $\R_{++}, \W_-$ and $\T_{----}$ are improved. On $\R_{--}$ it acts by:
\begin{equation}
\R_{--} \to \R_{--} - 2\partial_{--}\tilde{U},
\end{equation}
where the relation between $U$ and $\tilde{U}$ is as in \eqref{mir}, that is
\begin{align}
U 		&= u + i\theta^+\chi_+ + i\bar\theta^+\bar\chi_+ + \theta^+\bar\theta^+\partial_{++}v,\cr
\tilde{U}&=v + \theta^+\chi_+ - \bar\theta^+\bar\chi_+ - \theta^+\bar\theta^+\partial_{++}u.
\end{align}
For an improvement transformation of the R-multiplet to make sense, we have to assume that both $U$ and $\tilde{U}$ are local superfields. In view of the comment we made before, this restricts the class of allowed $U$. While for a general supercurrent multiplet the improvem	ent transformations were parametrized by an arbitrary local real superfield $U$, for the R-multiplet they are parametrized by such a local real superfields $U$ that $\tilde{U}$ is also local. Thus the R-multiplet allows a smaller class of improvements then a general supercurrent multiplet. This is not surprising after all. For the general supercurrent multiplet, only the stress-energy tensor and the supersymmetry currents are conserved, so improvements should only preserve their conservation. In the R-multiplet, on the other hand, we also have the conserved R-current, so preserving its conservation (and the R-charge value) restricts the class of allowed improvements.

In terms of component currents, the improvement transformation is:
\begin{align}
j_{++} &\to j_{++} + 2\partial_{++}v,\quad j_{--} \to j_{--} - 2\partial_{--}v,\cr
T_{++++} &\to T_{++++} + \partial_{++}^2 u,\quad T_{++--} \to T_{++--} -\partial_{++}\partial_{--}u,\quad T_{----} \to T_{----} + \partial_{--}^2 u,\cr
S_{+++} &\to S_{+++} + i\partial_{++}\chi_+,\quad S_{+--} \to S_{+--} -i\partial_{--}\chi_+,\cr
\bar{S}_{+++} &\to \bar{S}_{+++} - i\partial_{++}\bar\chi_+,\quad \bar{S}_{+--} \to \bar{S}_{+--} +i\partial_{--}\bar\chi_+.\cr
\end{align}
As expected, this transformation does not spoil conservation of any of these currents. It does not shift values of any conserved charges either. Also, it is easy to check that components $\tilde{T}_{++++}$ and $\tilde{T}_{++--}$ of the half-twisted stress-energy tensor are shifted by $\bar{Q}_+$-exact terms. On the other hand, $\tilde{T}_{----}$ is shifted by $\partial_{--}^2 (u+iv)$, which is, being the lowest component of chiral superfield $U+i\tilde{U}$, is $\bar{Q}_+$-closed but generally is not $\bar{Q}_+$-exact. Therefore, there is a family of possible holomorphic stress tensors in the $\bar{Q}_+$-cohomology, corresponding to different improvements.

Another ambiguity, namely shifting by the superspace current $\I_{\pm\pm}$, works in a straightforward way:
\begin{align}
\R_{++} &\to \R_{++} + \I_{++},\cr
\R_{--} &\to \R_{--} + \I_{--},\cr
\T_{----}&\to \T_{----} + \frac{1}{2}\partial_{--}\tilde\I_{--}.
\end{align}
If we denote the components of $\I_{\pm\pm}$ by:
\begin{equation}
\I_{\pm\pm} = i_{\pm\pm} - 2i\theta^+ I_{+\pm\pm} - 2i\bar\theta^+ \bar{I}_{+\pm\pm} -2\theta^+\bar\theta^+ H_{++\pm\pm},
\end{equation}
and introduce a local operator $h_{--}$ such that 
\begin{equation}
\partial_{++}h_{--}=H_{++--},
\end{equation}
then the shifting transformation in components works as:
\begin{align}
j_{\pm\pm} &\to j_{\pm\pm} + i_{\pm\pm},\cr
S_{+\pm\pm} &\to S_{+\pm\pm} + I_{+\pm\pm},\cr
\bar{S}_{+\pm\pm} &\to \bar{S}_{+\pm\pm} + \bar{I}_{+\pm\pm},\cr
T_{++\pm\pm} &\to T_{++\pm\pm} + H_{++\pm\pm},\cr
T_{----} &\to T_{----} - \partial_{--}h_{--}.
\end{align}
This ambiguity will naturally arise in a later discussion.
\subsubsection{OPE in the cohomology}\label{OPEcoh}
If we have two operators $\O_1$ and $\O_2$ representing nontrivial $\bar{Q}_+$-cohomology classes, we can consider their OPE. On very general grounds we have:
\begin{equation}
\O_1 (x^{++},x^{--}) \O_2(0,0) = \sum_{n,m} (x^{++})^n (x^{--})^m \O_{n,m}(0,0).
\end{equation}
Now recall that the operator $\partial_{++}$ acts trivially in the cohomology, that is if $\O_1$ is $\bar{Q}_+$-closed, then $\partial_{++}\O_1$ is $\bar{Q}_+$-exact, and thus so is $\partial_{++}\O_1 (x^{++},x^{--}) \O_2(0,0)$. Acting with $\partial_{++}$ on the right-hand side then gives a $\bar{Q}_+$-exact answer, that is:
\begin{equation}
\sum_{n,m} n (x^{++})^{n-1} (x^{--})^m \O_{n,m}(0,0) = [\bar{Q}_+, \dots].
\end{equation}
This implies that all terms except those with $n=0$ are $\bar{Q}_+$-exact. If the cohomology classes represented by $\O_i$ have scaling dimensions $h_i$, we can then write:
\begin{equation}
\O_1 (x^{++},x^{--}) \O_2(0,0) = \sum_k \frac{1}{(x^{--})^{h_1+h_2-h_k}} \O_k(0,0) + [\bar{Q}_+,\dots].
\end{equation}
Note also that, since in the cohomology we have left-movers only, scaling dimensions and spins coincide\footnote{Even if the operator representing the cohomology class in the full theory is not left-moving, the class it represents is left-moving. Since Lorenz-invariance of the full theory induces Lorenz-invariance in the cohomology, one indeed can use the argument made in the text. }. This, in particular, implies an obvious conclusion that no dimensionful constants can appear in the OPE of the cohomology classes. Any dimensionful constant will have non-trivial dimension but trivial spin, and therefore its appearance will either break scaling or Lorentz-invariance of the OPE. Indeed, if we have some dimensionful parameter $\mu$, then in the expression: 
\begin{equation}
\O_1 (x^{++},x^{--}) \O_2(0,0) = \sum_k \frac{\mu^{p}}{(x^{--})^\Delta} \O_k(0,0) + [\bar{Q}_+,\dots],
\end{equation}
scaling invariance implies $\Delta= h_1 + h_2 - h_k - h(\mu)p$, where $h(\mu)$ is the dimension of $\mu$, while Lorentz invariance implies $\Delta = h_1 + h_2 - h_k$. This is possible only for $p=0$, that is $\mu$ should not be there.

All dependence on dimensionful coupling constants of the original supersymmetric theory will therefore be hidden in the $\bar{Q}_+$-exact term. This simple observation will be helpful later. It will imply that one can turn off all dimensionful couplings for the OPE computation. In the models we are going to study this will mean that it is enough to compute OPE in the free theory.

\subsection{Chiral algebras of superconformal theories}\label{SCFT}
For superconformal theories, the $\N=(0,2), d=2$ super-Poincare algebra of symmetries is enhanced to ${\rm Vir}\oplus \t{\rm SVir}$, where ${\rm Vir}$ denotes the left-handed Virasoro algebra (generated by the holomorphic stress-energy tensor) and the $\t{\rm SVir}$ denotes the right-handed $\N=2$ super-Virasoro algebra (generated by the corresponding anti-holomorphic currents). The left-handed algebra might be enlarged to the super-Virasoro as well (or even some larger W-algebra) if we have more symmetries on the left, but it graded-commutes with the $\N=2$ Virasoro on the right in any case. 

Let us restrict to the NS sector of the $\t{\rm SVir}$.
The operators $\bar{Q}_+$ and $Q_+$ can be identified as $\tilde{G}^+_{-1/2}$ and $\tilde{G}^-_{-1/2}$ respectively -- two of the fermionic generators of $\t{\rm SVir}$ (we put tildes on $\t{\rm SVir}$ and on its generators to emphasize that this is an anti-holomorphic algebra). In a conformal case, we have the radial quantization Hilbert space $\H$, and we assume that it has an inner product, such that $\tilde{G}^-_{1/2}=\left(\tilde{G}^+_{-1/2}\right)^\dagger$ is a special supersymmetry generator. Part of the super-Virasoro algebra relations are:
\begin{align}
\label{svir}
\{\tilde{G}^-_{-1/2}, \tilde{G}^+_{-1/2}\}&=2\tilde{L}_{-1},\cr
\{\tilde{G}^+_{-1/2}, (\tilde{G}^+_{-1/2})^\dagger\}\equiv\{\tilde{G}^+_{-1/2}, \tilde{G}^-_{1/2}\}&=2\tilde{L}_0-\tilde{J}_0.
\end{align}
Recall that in conformal case we have a state-operator correspondence. Therefore, instead of computing the operator cohomology, we can equivalently ask for the cohomology of $\tilde{G}^+_{-1/2}$ acting on the Hilbert space $\H$. The second equation in \eqref{svir} shows that, by the standard Hodge theory argument, this cohomology can be identified with the kernel of $2\tilde{L}_0 - \tilde{J}_0$. Also, in a unitary theory, it shows that $2\tilde{L}_0 -\tilde{J}_0\ge 0$.

Now, every state in the Hilbert space is built by acting with $\tilde{L}_{-n}, \tilde{J}_{-n}, \tilde{G}^+_{-\alpha}, \tilde{G}^-_{-\alpha}, n,\alpha>0$ on a superconformal primary state. It is easy to see that all these operators except $\tilde{G}^+_{-1/2}$ increase the eigenvalue of $2\tilde{L}_0 - \tilde{J}_0$, while $\tilde{G}^+_{-1/2}$ does not change it. Therefore, if the primary state has $2\tilde{L}_0 - \tilde{J}_0>0$, then all states in its superconformal family have $2\t{L}_0 - \t{J}_0 >0$ and thus do not contribute to the cohomology. On the other hand, if some primary state $|\Delta\rangle$ has zero eigenvalue of $2\t{L}_0 - \t{J}_0$, then so does $\t{G}^+_{-1/2}|\Delta\rangle$, while other states in the same conformal family have $2\t{L}_0 - \t{J}_0 >0$. But $(2\t{L}_0 - \t{J}_0)|\Delta\rangle=0$ and the second equation of \eqref{svir} imply that $\t{G}^+_{-1/2}|\Delta\rangle=0$. Therefore, in such a case there is just one non-trivial state in the superconformal family which contributes to the cohomology -- the primary state itself. This way we prove that in the NS sector of a unitary $\N=(0,2)$ superconformal theory there is an isomorphism:
\begin{align}
\label{ring02}
H (\H, \tilde{G}^+_{-1/2}) \simeq \{{\rm Primaries\ of\ \t{SVir}\ with\ } 2\t{L}_0 - \t{J}_0=0\}\cr
=\{|\psi\rangle\in\H: \t{L}_{n}|\psi\rangle=\t{J}_{n}|\psi\rangle=\t{G}^+_{\alpha-1}|\psi\rangle=\t{G}^-_{\alpha}|\psi\rangle=(2\t{L}_0 - \t{J}_0)|\psi\rangle=0, n,\alpha>0\}.\cr
\end{align}
Notice that these are what is usually called the chiral primaries with respect to ${\rm \t{SVir}}$. In fact, this is essentially the construction of \cite{LVW} applied to $\N=(0,2)$ theories. In the $\N=(2,2)$ case, \cite{LVW} describe the chiral ring of the $\N=(2,2)$ model by studying the set of (anti)chiral primaries both with respect to the left- and the right-moving super-Virasoro algebras. For the $\N=(0,2)$ theories, we have in \eqref{ring02} only the chiral primary condition with respect to the right-moving super-Virasoro algebra. For that reason, the object we get is not just the chiral ring: it involves holomorphic OPEs as part of its structure and is usually referred to as the W-algebra, or also chiral algebra.

Another remark is that for $\N=(2,2)$ theories, the chiral algebra that we study encodes the $(c,c)$ and $(a,c)$ rings of \cite{LVW} as a part of its structure. Indeed, by considering the subspace of $H(\H, \t{G}^+_{-1/2})$ annihilated by $2L_0-J_0$, where $L_0$ and $J_0$ are from the left-moving ${\rm SVir}$ algebra, we get the space $\{|\psi\rangle\in\H:\, (2L_0-J_0)|\psi\rangle=(2\t{L}_0-\t{J}_0)|\psi\rangle=0\}$, which is the space of chiral primaries with respect to both ${\rm SVir}$ and ${\rm \t{SVir}}$, and therefore gives rise to the $(c,c)$ ring under the OPE. Analogously, picking the subspace annihilated by $L_0 + J_0$, we get the $(a,c)$ ring.

One consequence of this is that in $\N=(2,2)$ theories, the (anti)chiral primaries, which form the $(c,c)$ or $(a,c)$ rings of the theory, always show up in the chiral algebra as primaries of the left-moving ${\rm \t{SVir}}$. In the simplest cases they will generate the whole chiral algebra, but as we will see later, there might be other primary operators in the algebra, which are not simply elements of the $(c,c)$ or $(a,c)$ ring.

\subsection{The operator cohomology and the superspace}
\subsubsection{Classical and quantum observables}
In the models we are going to study later in this paper, the chiral algebra will turn out to be tree-level exact. As we will argue, no loop corrections will contribute to the cohomology. However, despite our usual intuition that ``tree level'' means ``classical'', it is important to understand that the quantum chiral algebra in the $\bar{Q}_+$-cohomology is not the same as the classical one. The distinction comes from the way we multiply operators.

In classical field theory, to multiply fields we use the usual point-wise multiplication of functions on space-time. In quantum theory, even at the tree level, we should subtract singularities which appear when different operators collide, which for example gives the usual notion of normal ordering in CFT.

It might happen (and it will happen in concrete examples) that the classical composite operator is $\bar{Q}_+$-closed, but the singular part we need to subtract to define the quantum operator is not $\bar{Q}_+$-closed. This subtlety should be taken into account when computing the chiral algebra of the theory. But still, as a step in this direction, it is useful to understand the structure of the classical chiral algebra first.

\subsubsection{Classical observables and the cohomology}
Let us introduce the space of classical observables $\F$ and the space of classical superobservables $\hF$. We will sometimes refer to a generic field as $\phi$ and to a generic superfield as $\Phi$. Both of these spaces classically carry the structures of supercommutative algebras. 

\textbf{Definition 2.1: } $\F$ is a supercommutative algebra of polynomials of fields $\phi$ and their derivatives $\partial_{--}^n\partial_{++}^m\phi$ whose coefficients are analytic functions on a space-time, modulo classical equations of motion. In other words, $\F=C_\omega(M)[\dots, \phi, \partial_{--}^n\partial_{++}^m\phi,\dots]/\I$, where $C_\omega(M)$ denotes analytic functions on $M$, and $\I$ denotes an ideal generated by the equations of motion and all their derivatives.

If the classical equations of motion do not depend on space-time coordinates explicitly (only through the coordinate-dependence of the generating fields), we can introduce:

\textbf{Definition 2.1': } $\F_0$ is a subalgebra of $\F$ of observables which do not depend on a space-time point explicitly. In other words, it is generated by the same fields and their derivatives as $\F$ (and also modulo equations of motion), but the coefficients are taken to be just complex numbers rather than functions.

There are straightforward superspace analogs of these:

\textbf{Definition 2.2: } $\hF$ is a supercommutative algebra of polynomials of superfields $\Phi$ and their bosonic and super-derivatives $\partial_{--}^n\partial_{++}^mD_+^p \bar{D}_+^q\Phi$ whose coefficients are analytic functions on superspace, modulo classical superspace equations of motion.

If the superspace equations of motion do not include any explicit dependence on a superspace point, i.e. if they have the form of a polynomial of generating fields $\partial_{--}^n\partial_{++}^mD_+^p \bar{D}_+^q\Phi$ with complex coefficients, we again can define a subalgebra:

\textbf{Definition 2.2': } $\hF_0$ is a subalgebra of $\hF$ of superobservables which do not depend on a space-time point explicitly. In other words, it is generated by the same superfields and their derivatives as $\hF$ (and also modulo superspace equations of motion), but the coefficients are taken to be just complex numbers rather then functions.

Our goal is to compute the cohomology of $\bar{Q}_+$ acting on $\F$ in the situation when the equations of motion do not depend on the superspace point explicitly. The first observation is that the operator $\bar{Q}_+$ only acts on the generating fields of the algebra $\F$, it does not act on the c-number functions which can possibly multiply these fields. This means that it is enough to compute the cohomology of $\bar{Q}_+$ acting on $\F_0$. To be more rigorous, we can introduce operators of multiplication by $x^\mu$ called $m(x^\mu)$:
\begin{equation}
\forall \mathcal{O}\in\F,\ m(x^\mu)\mathcal{O} = x^\mu\mathcal{O},
\end{equation}
and notice that they commute with $\bar{Q}_+$. Then we can introduce a bigrading on $\F$ by saying that an explicit factor of $(x^0)^n(x^1)^m$ has degree $(n,m)$. After this it becomes obvious that
\begin{equation}
H(\F)\simeq \overline{\bigoplus_{n,m\ge 0} H^{n,m}(\F)},
\end{equation}
where the bar over the right hand side means that we should actually consider a completion of this space with respect to some norm, because we have to allow infinte sums (series) to account for the possibility of having analytic functions as coefficients.

As we mentioned, $\bar{Q}_+$ does not act in any way on $x^\mu$, and because of that:
\begin{equation}
H^{n,m}(\F)\simeq H(\F_0).
\end{equation}
Therefore, from now on we will only study the cohomology in $\F_0$, which of course only makes sense when the equations of motion do not depend on the superspace point explicitly.
\subsubsection{The cohomology of $\bar{Q}_+$ in $\F_0$ and of $\bar{D}_+$ in $\hF_0$}
Take an arbitrary $\mathcal{A}\in\hF_0$. $\mathcal{A}$ is some general superfield, and it can be expanded into components with respect to the Grassmann coordinates. The most basic property it satisfies is that the supersymmetry transformations of its components are encoded in the way differential operators $\Q_+$ and $\bar\Q_+$ act on it. This follows simply from the fact that this holds for the generating superfields from which $\mathcal{A}$ is constructed and the fact that we do not allow explicit dependence on the superspace coordinates in the algebra $\hF_0$. So we have:
\begin{equation}
[Q_+, \mathcal{A}]= \Q_+\mathcal{A},
\end{equation}
and the same for $\bar{Q}_+$. Supersymmetry relates all components of $\mathcal{A}$ and it is straightforward to see that:

\textbf{Proposition 2.1:} If the lowest component $\mathcal{A}\big|$ of the superfield $\mathcal{A}\in\hF_0$ vanishes, then $\mathcal{A}=0$.

The algebras $\F_0$ and $\hF_0$ are related in an obvious way: any element of $\F_0$ can be found as a component of some superfield in $\hF_0$. In particular, we can always find a superfield $\mathcal{A}$ which contains a given element $a\in\F_0$ as its lowest component. Moreover, supersymmetry defines this $\mathcal{A}$ uniquely, so:

\textbf{Proposition 2.2:} For any $a\in \F_0$ there exists a unique $\mathcal{A}\in\hF_0$ such that $a=\mathcal{A}\big|$.

The problem which we are addressing is to find the cohomology of $\bar{Q}_+$ in $\F_0$. That is, the classes of fields $a\in\F_0$ which satisfy $[\bar{Q}_+,a]=0$, modulo those $a$ for which $a=[\bar{Q}_+,b], b\in\F_0$. Now from the Proposition 2, we know that there exist $\mathcal{A}, \mathcal{B} \in\hF_0$, such that $a=\mathcal{A}|$ and $b=\mathcal{B}|$. The equation $[\bar{Q}_+,a]=0$ implies then $\bar\Q_+\mathcal{A}|=0$. 

There is a small subtlety here which shows why it is correct to look for the cohomology of $\bar{D}_+$ rather than $\bar\Q_+$: $\bar{D}_+$ acts on $\hF_0$ by definition, while $\bar\Q_+=\bar{D}_+ + 2i\theta^+\partial_+$ does not, as it introduces an explicit dependence on $\theta$ (therefore $\bar\Q_+$ acts from $\hF_0$ to a bigger space $\hF$). However, we can write: $\bar{D}_+\mathcal{A}|=\bar\Q_+\mathcal{A}|=0$. But $\bar{D}_+\mathcal{A}\in\hF_0$, so we can apply Proposition 1 and conclude that $\bar{D}_+\mathcal{A}=0$. Analogously $a=[\bar{Q}_+,b]$ implies $\mathcal{A}=\bar{D}_+\mathcal{B}$. This proves the

\textbf{Proposition 2.3:} The cohomology of $\bar{Q}_+$ in $\F_0$ (denoted $H(\F_0)$) is isomorphic to the cohomology of $\bar{D}_+$ in $\hF_0$ (denoted $H(\hF_0)$). The isomorphism $H(\hF_0)\to H(\F_0)$ is defined by taking the lowest component of the superfield.

This proposition shows why in the rest of the paper we are going to study the cohomology of $\bar{D}_+$.
\section{Landau-Ginzburg models}
The $\N=(0,2)$ Landau-Ginzburg (LG) model is described by a set of chiral superfields $\Phi^i, i=1..n$ and Fermi superfields $\Lambda^a, a=1..m$. The action is (we assume summation over repeated indices, even if they both appear upstairs or downstairs; sometimes we will write the sum sign explicitly to avoid possible confusion):
\begin{align}
\label{LGaction}
S=\frac{1}{\pi}\int\d^2x\d^2\theta \left\{ -\frac{i}{2}\bar\Phi^i \partial_{--}\Phi^i -\frac{1}{2}\bar\Lambda^a\Lambda^a \right\} + \frac{1}{\pi}\int \d^2x\d\theta^+\Lambda^a J_a(\Phi)|_{\bar\theta^+=0} + h.c.,
\end{align}
where in general $\bar{D}_+\Lambda^a = E^a(\Phi)$ and $\sum_a E^a(\Phi) J_a(\Phi)=0$. The classical superspace equations of motion are:
\begin{align}
\label{LG_EOM}
\bar{D}_+\partial_{--}\bar\Phi^i &= i\bar\Lambda^a \frac{\partial E^a}{\partial\Phi^i} -2i \Lambda^a \frac{\partial J_a}{\partial\Phi^i},\cr
D_+\partial_{--}\Phi^i &= -i\Lambda^a \frac{\partial\bar E^a}{\partial\bar\Phi^i} +2i \bar\Lambda^a \frac{\partial\bar J_a}{\partial\bar\Phi^i},\cr
\bar{D}_+\bar\Lambda^a &= -2J_a(\Phi),\cr
D_+\Lambda^a &= 2\bar{J}_a(\bar\Phi).
\end{align}
The supersymmetry currents of this theory are:
\begin{align}
\label{sucurr}
S_{+++}&=\frac{i}{2}\psi^i_+\partial_{++}\bar\phi^i,\quad \bar{S}_{+++} = -\frac{i}{2}\bar\psi^i_+\partial_{++}\phi^i,\cr
S_{+--}&=\frac{i}{2}\lambda^a\bar{E}^a(\bar\phi) - i\bar\lambda^a\bar{J}_a(\bar\phi),\quad \bar{S}_{+--}=i\lambda^a J_a(\phi) - \frac{i}{2}\bar\lambda^a E^a(\phi).
\end{align}
It is not hard to find a superfield $\S_{++}$ such that $S_{+++}=\frac{i}{2}D_+\S_{++}|$ and $\bar{S}_{+++}=-\frac{i}{2}\bar{D}_+\S_{++}|$:
\begin{align}
\S_{++}= \frac{1}{2}D_+\Phi^i \bar{D}_+\bar\Phi^i.
\end{align}
If we also introduce:
\begin{align}
\W_- &= \frac{i}{2}\bar\Lambda^a E^a - i\Lambda^a J_a,\cr
\T_{----}&=\partial_{--}\Phi^i\partial_{--}\bar\Phi^i +\frac{i}{2}\Lambda^a\partial_{--}\bar\Lambda^a -\frac{i}{2}\partial_{--}\Lambda^a \bar\Lambda^a,
\end{align}
then we find that:
\begin{align}
\label{sucurrmult}
\partial_{--}\S_{++} &=D_+\W_- -\bar{D}_+\bar\W_-,\cr
\bar{D}_+\T_{----} &=\partial_{--}\W_-,\cr
\bar{D}_+\W_- &=0.
\end{align}
We see that these are precisely the relations \eqref{constr} of the $\N=(0,2)$ $d=2$ supercurrent, and moreover, the component expansions of $\S_{++}$, $\W_-$ and $\T_{----}$, written as in \eqref{components}, include the supersymmetry currents \eqref{sucurr}. Therefore, we have described the supercurrent multiplet of the theory \eqref{LGaction}. In a generic situation, it is not an R-multiplet, because there are no R-symmetries.

The algebra $\hF_0$ is a supercommutative algebra freely generated by superfields $\Phi^i, \bar\Phi^i, \Lambda^a, \bar\Lambda^a$ and their derivatives (with respect to $\partial_{--}$, $\partial_{++}$, $D_{+}$ and $\bar{D}_{+}$ applied arbitrary number of times) modulo the relations. The relations are: the ones that follow from $\{D_+,\bar{D}_+\}=2i\partial_{++}$ and $D_+^2=\bar{D}_+^2=0$, the chirality conditions $\bar{D}_+\Phi^i=0$, $\bar{D}_+\Lambda^a=E^a(\Phi)$ and the superspace equations of motion as written above. All differential corollaries of the relations should also be included as relations of course.

It is not too hard to find a set of independent generators $\mathcal{G}$, so that all the relations will be taken into account and we will have simply $\hF_0\simeq \C[\mathcal{G}]$, a polynomial algebra generated by those generators. 

We will now find this $\mathcal{G}$. First of all, due to the chirality conditions, $\Phi^i$ can only appear with the $D_+$ derivative (moreover, with at most one, because $D_+^2=0$), and $\bar\Phi^i$ -- with $\bar{D}_+$. The chirality condition for $\Lambda^a$ allows to replace $\bar{D}_+\Lambda^a$ by $E^a(\Phi)$, while the equation of motion $D_+\Lambda^a=-2\bar{J}_a(\bar\Phi)$ allows to replace $D_+\Lambda^a$ by an expression without derivatives. Therefore it is enough to consider only bosonic derivatives acting on $\Lambda^a$. However the simple relation:
\begin{equation}
2i\partial_{++}\Lambda^a = \{D_+,\bar{D}_+\}\Lambda^a = D_+\bar{D}_+\Lambda^a + \bar{D}_+ D_+\Lambda^a = D_+E^a(\Phi) +2\bar{D}_+\bar{J}_a(\bar\Phi)
\end{equation}
shows that $\partial_{++}$ derivatives acting on Fermi superfields can also be removed. Therefore, in the generating set $\mathcal{G}$, it is enough to include only $\partial_{--}^n\Lambda^a$ and $\partial_{--}^n\bar\Lambda^a$, with $n\ge 0$, and the appropriate derivatives of bosonic chiral superfields. By appropriate derivatives of bosonic chiral superfields we mean the following. First, we need to include $\partial_{--}^n\Phi^i$ and $\partial_{--}^n\bar\Phi^i$ with $n\ge0$. $D_+\Phi^i$ and $\bar{D}_+\bar\Phi^i$ should also be included, but there is no need to include expressions like $\bar{D}_+\partial_{--}^n\Phi^i$, because, as equations of motion for $\Phi^i$ show, $D_+\partial_{--}\Phi^i$ and $\bar{D}_+\partial_{--}\bar\Phi^i$ can be replaced by expressions without derivatives. Expressions like $\partial_{++}^n\Phi^i$, $D_+\partial_{++}^n\Phi^i$ and their complex conjugates have to be included, they cannot be reduced to expressions without derivatives. Finally, there is no need to include both $\partial_{++}$ and $\partial_{--}$ derivatives because of:
\begin{equation}
2i\partial_{++}\partial_{--}\bar\Phi^i=D_+\bar{D}_+\partial_{--}\bar\Phi^i=D_+\left(i\bar\Lambda^a \frac{\partial E^a}{\partial\Phi^i} -2i \Lambda^a \frac{\partial J_a}{\partial\Phi^i}\right).
\end{equation}
So, to summarize, we write the generating set explicitly:
\begin{align}
\mathcal{G}=\big\{\partial_{--}^n\Phi^i, \partial_{--}^n\bar\Phi^i, \partial_{++}^n\Phi^i, D_+\partial_{++}^n\Phi^i, \partial_{++}^n\bar\Phi^i, \bar{D}_+\partial_{++}^n\bar\Phi^i, \partial_{--}^n\Lambda^a, \partial_{--}^n\bar\Lambda^a, n\ge 0 \big\}.
\end{align}
To emphasize once again, we claim that:
\begin{equation}
\hF_0 \simeq \C[\mathcal{G}].
\end{equation}
Using the relations satisfied by the fields, it is not hard to describe the action of $\bar{D}_+$ in terms of the generators in $\mathcal{G}$. We have:
\begin{align}
\bar{D}_+(\partial_{--}^n\Phi^i)&=0,\cr
\bar{D}_+(\partial_{--}^n\bar\Phi^i)&= \partial_{--}^{n-1}\left(i\bar\Lambda^a \frac{\partial E^a}{\partial\Phi^i} -2i \Lambda^a \frac{\partial J_a}{\partial\Phi^i} \right),\cr
\bar{D}_+(\partial_{++}^n\Phi^i)&=0,\quad \bar{D}_+ (D_+\partial_{++}^n\Phi^i)=2i\partial_{++}^{n+1}\Phi^i,\cr
\bar{D}_+(\partial_{++}^n\bar\Phi^i)&= \bar{D}_+\partial_{++}^n\bar\Phi^i,\quad \bar{D}_+(\bar{D}_+\partial_{++}^n\bar\Phi^i)=0,\cr
\bar{D}_+(\partial_{--}^n\Lambda^a)&=\partial_{--}^n E^a(\Phi),\cr
\bar{D}_+(\partial_{--}^n\bar\Lambda^a)&=-2\partial_{--}^nJ_a(\Phi).
\end{align}
From these formulas we can guess that polynomials of $\partial_{--}^n\Phi^i$ should be in the cohomology. However, we need some extra assumptions about $E^a$ and $J_a$ in order to move further.
\subsection{Quasihomogeneous case}
As we have already learned, it is interesting to consider the case when the theory has an R-symmetry. In such a case, we expect to have a stress-energy tensor in the cohomology. It is not hard to check that the following transformation:
\begin{align}
\theta^+ &\to e^{-i\epsilon}\theta^+,\cr
\Phi^i &\to e^{-i\epsilon\alpha_i}\Phi^i,\cr
\Lambda^a &\to e^{-i\epsilon\t\alpha_a}\Lambda^a
\end{align}
is a symmetry of the classical action if and only if the following quasihomogeneity conditions are satisfied:
\begin{align}
\label{quasih}
\t\alpha_a J_a + \sum_i \alpha_i\Phi^i \frac{\partial J_a}{\partial\Phi^i} &=J_a,\cr
-\t\alpha_a E^a +\sum_i \alpha_i\Phi^i\frac{\partial E^a}{\partial\Phi^i} &= E^a,
\end{align}
where $\alpha_i$ and $\t\alpha_a$ are real numbers. It is a matter of a standard calculation to find the real conserved current $j_{\pm\pm}$ for this R-symmetry. It is then straightforward to write a superfield which has it as the lowest component. The answer is:
\begin{align}
\R_{++}&=-\frac{i}{2}\sum_i \alpha_i\left(\Phi^i\partial_{++}\bar\Phi^i - \bar\Phi^i\partial_{++}\Phi^i\right) +\frac{1}{2}\sum_i(1-\alpha_i)D_+\Phi^i\bar{D}_+\bar\Phi^i\cr
\R_{--}&=-\frac{i}{2}\sum_i\alpha_i\left(\Phi^i\partial_{--}\bar\Phi^i-\bar\Phi^i\partial_{--}\Phi^i\right)- \sum_a\t\alpha_a\Lambda^a\bar\Lambda^a.
\end{align}
It is not a coincidence that we called it $\R$. In fact, one can check that the equations of motion imply $\partial_{--}\R_{++}+\partial_{++}\R_{--}=0$. Therefore, higher components of $\R$ are also conserved currents. This is the supercurrent multiplet discussed before provided we can find another real superfield $\Y_{----}$ (which has possibly improved stress-energy tensor $T_{----}$ as its lowest component) satisfying the required constraints. As one can check from \eqref{constr}, the condition\footnote{The fact that this $\Y_{----}$ together with $\Y_{++++}=\frac{i}{4}[D_+,\bar{D}_+]\J_{++}$ and $\Y_{++--}=\frac{i}{4}[D_+,\bar{D}_+]\J_{--}$ form a conserved superspace current then follows automatically.} on $\Y_{----}$ is:
\begin{align}
\bar{D}_+\Y_{----} &= \frac{i}{2}\bar{D}_+\partial_{--}\R_{--}.
\end{align}
Note that $D_+\Y_{----} = -\frac{i}{2}D_+\partial_{--}\R_{--}$ is then satisfied automatically. This defines $\Y_{----}$ uniquely up to an arbitrary function of $x^-$, as $\Y_{----}\to \Y_{----}+ f(x^-)$ preserves the above constraints. A simple computation allows to find a real superfield such that it satisfies these constraints:
\begin{equation}
\Y_{----}=\sum_i\left[\partial_{--}\Phi^i\partial_{--}\bar\Phi^i - \frac{\alpha_i}{4}\partial_{--}^2(\bar\Phi^i\Phi^i)\right] +\sum_a\left[\frac{i}{2}\Lambda^a\partial_{--}\bar\Lambda^a - \frac{i}{2}\partial_{--}\Lambda^a\bar\Lambda^a\right].
\end{equation}
Therefore, we actually have a supercurrent multiplet described by $\R_{++}$, $\R_{--}$ and $\Y_{----}$, which is, moreover, an R-multiplet in the terminology of \cite{DS}, as reviewed in Section \ref{sec_supcurr}. The $\bar{Q}_+$-cohomology class represented by the twisted stress-energy tensor $\tilde{T}_{----}$ from \eqref{tw} promotes to the $\bar{D}_+$-cohomology class represented by the superfield:
\begin{align}
\label{stressT}
\tilde\Y&=\Y_{----}-\frac{i}{2}\partial_{--}\J_{--}\cr
&=\sum_i\left[\partial_{--}\Phi^i\partial_{--}\bar\Phi^i-\frac{\alpha_i}{2}\partial_{--}(\Phi^i\partial_{--}\bar\Phi^i)\right] +\sum_a\left[\frac{i}{2}\Lambda^a\partial_{--}\bar\Lambda^a - \frac{i}{2}\partial_{--}\Lambda^a\bar\Lambda^a +\frac{i\t\alpha_a}{2}\partial_{--}(\Lambda^a\bar\Lambda^a)\right].\cr
\end{align}
This is precisely the stress-energy tensor in the cohomology as found in \cite{SW}. At first sight, one could think that this is the end of the story. However, there are some subtleties here, which we will now discuss.

First of all, how is this R-multiplet related to the more general supercurrent multiplet which we found in \eqref{sucurrmult}? The answer is simple. If we also define 
\begin{equation}
\mathcal{V}_-= \frac{i}{2}\bar{D}_+\R_{--},
\end{equation}
then $\R_{++}$, $\mathcal{V}_-$ and $\Y_{----}$ form a supercurrent multiplet related to $\S_{++}$, $\W_-$ and $\T_{----}$ by the improvement transformation:
\begin{align}
\R_{++} &= \S_{++} + [D_+,\bar{D}_+]U,\cr
\mathcal{V}_- &= \W_- + \partial_{--}\bar{D}_+ U,\cr
\Y_{----} &= \T_{----} + \partial_{--}^2U,\cr
U&=-\sum_i \frac{\alpha_i}{4}\Phi^i\bar\Phi^i.
\end{align}
Note that the superfield $U$ cannot be represented as a real part of some \emph{local} chiral superfield. Therefore this is an example of the improvement transformation allowed for the general supercurrent multiplet but not allowed for the R-multiplet. As we will see momentarily, there might exist several R-multiplets which are not equivalent to each other as R-mulitplets (cannot be related to each other by the R-multiplet improvements), but they all are related to the same supercurrent multiplet $\S_{++}, \W_-, \T_{----}$ by the more general improvement described above.

So now we will discuss the possibility of having several inequivalent R-multiplets. Note that the quasihomogeneity conditions \eqref{quasih} might have more than one solution. This corresponds to having an extra flavor $U(1)$ symmetry, which can then mix with the R-symmetry to give another solution of \eqref{quasih} (in terms of current, this means to replace the R-symmetry current $j_R$ by $j_R+j_F$, where $j_F$ is a Flavor symmetry current).

The flavor symmetry does not rotate the thetas, so it acts just as:
\begin{align}
\label{flavor}
\Phi^i &\to e^{-i\epsilon q_i}\Phi^i,\cr
\Lambda^a &\to e^{-i\epsilon\t q_a}\Lambda^a.
\end{align}
The condition that this is a symmetry of the classical action is:
\begin{align}
\label{quasih2}
\t q_a J_a + \sum_i q_i\Phi^i \frac{\partial J_a}{\partial\Phi^i} &=0,\cr
-\t q_a E^a +\sum_i q_i\Phi^i\frac{\partial E^a}{\partial\Phi^i} &= 0.
\end{align}
We can see now that if $\{\alpha_i, \t\alpha_a\}$ is some solution of \eqref{quasih} and $\{q_i, \t q_a\}$ is some solution of \eqref{quasih2}, then $\{\alpha_i+q_i, \t\alpha_a+\t q_a\}$ is another solution of \eqref{quasih}. This is actually the ambiguity of the supercurrent multiplet which we were discussing before. In case we have extra superspace currents, the basic supercurrent mutiplet $\S_{++}, \W_-, \T_{----}$ can be shifted. Let us belabor this point somewhat further.

One can compute the current corresponding to the flavor symmetry \eqref{flavor} and find the real superfield which contains it as the lowest component:
\begin{align}
\I_{--} &= -\sum_a \t q_a \Lambda^a\bar\Lambda^a -\frac{i}{2}\sum_i q_i\left(\Phi^i\partial_{--}\bar\Phi^i-\bar\Phi^i\partial_{--}\Phi^i\right),\cr
\I_{++}&=-\frac{1}{2}\sum_i q_i D_+\Phi^i\bar{D}_+\bar\Phi^i + \frac{i}{2}\sum_i q_i\left(\bar\Phi^i\partial_{++}\Phi^i - \Phi^i\partial_{++}\bar\Phi^i\right).
\end{align}
On shell these are conserved at the level of superfields:
\begin{equation}
\partial_{++}\I_{--} + \partial_{--}\I_{++}=0.
\end{equation}
One can do a small computation to check that the following superfield:
\begin{equation}
\F_{--} = -\sum_a\t q_a\Lambda^a\bar\Lambda^a -i\sum_i q_i\Phi^i\partial_{--}\bar\Phi^i
\end{equation}
is chiral on-shell, i.e. it satisfies $\bar{D}_+\F_{--}=0$ provided the equations of motion hold. In particular, it means that this $\F_{--}$ gives rise to the left-moving $U(1)$ current in the cohomology. But it is also true that:
\begin{equation}
{\rm Re}\, \F_{--}= \I_{--}.
\end{equation}
Therefore, there exists a local expression for the superfield $\tilde\I_{--}$:
\begin{equation}
\tilde\I_{--} = {\rm Im}\, \F_{--} = -\frac{1}{2}\sum_i q_i\partial_{--} (\Phi^i\bar\Phi^i).
\end{equation}
So, according to the general discussion from the Section \ref{sec_supcurr}, we can shift the R-multiplet using this $\I_{\pm\pm}$. Recall that the shift is:
\begin{align}
\R_{++} &\to \R_{++} + \I_{++},\cr
\R_{--} &\to \R_{--} + \I_{--},\cr
\Y_{----} &\to \Y_{----} + \frac{1}{2}\partial_{--}\tilde\I_{--}.
\end{align}
For the $\bar{D}_+$-closed element $\tilde\Y_{----}=\Y_{----} - \frac{i}{2}\J_{--}$, we have:
\begin{equation}
\tilde\Y_{----} \to \tilde\Y_{----}-\frac{i}{2}\partial_{--}(\I_{--} + i\tilde\I_{--})=\tilde\Y_{----} - \frac{i}{2}\partial_{--}\F_{--}.
\end{equation}
So the cohomology class  $[\tilde\Y_{----}]$ gets shifted by $-\frac{i}{2}[\partial_{--}\F_{--}]$.

Let us summarize. We have the family of R-current multiplets generated by shifts using the superspace current $\I_{\pm\pm}$. In the cohomology this corresponds to having an extra left-moving $U(1)$ current $[\F_{--}]$ generating an ambiguity of the stress-energy tensor in the cohomology, as we can do shifts of the cohomology class $[\tilde\Y_{----}]$ by $[\partial_{--}\F_{--}]$.

But the conformal theory to which our LG model flows in the IR supposedly should have a unique stress-energy tensor, which thus gives a preferred stress-energy tensor for the chiral algebra in the $\bar{Q}_+$-cohomology. One can ask a natural question: which of the R-current multiplets above corresponds to the true stress-energy tensor of the theory in the IR? The answer is simple: the correct stress-energy tensor is the one, for which the $U(1)$ current $[\F_{--}]$ is a primary operator in the cohomology, at least when it is possible to make it primary (we will discuss this point later). It is clear that this corresponds to extremizing the central charge of the corresponding Virasiro algebra (see the next subsection). To turn this statement into a criteria for picking the unique solution $(\alpha_i, \t\alpha_a)$ of \eqref{quasih}, we need to understand first how to compute the operator product expansions (OPE) in the cohomology.

\subsubsection{The OPE in the cohomology}\label{OPcoh}
The component action of the model that we study is:
\begin{equation}
S = S_D + S_F,
\end{equation}
where the D-term action is:
\begin{align}
S_D =\frac{1}{\pi}\int\d^2x \Big(-\partial_{--}\bar\phi^i\partial_{++}\phi^i -\frac{i}{2}\bar\psi^i_+\partial_{--}\psi^i_+ - i\bar\lambda^a\partial_{++}\lambda^a - \frac{1}{2}G^a\bar{G}^a\cr 
+ \frac{i}{2}\partial_i E^a(\phi) \bar\lambda^a \psi^i_+ - \frac{i}{2}\partial_i\bar{E}^a(\bar\phi)\bar\psi^i_+\lambda^a + \frac{1}{2}E^a(\phi) \bar{E}^a(\bar\phi) \Big),
\end{align}
and the F-term is:
\begin{equation}
S_F = \frac{1}{\pi}\int\d^2x \left( G^a J_a(\phi) + \bar{G}^a\bar{J}_a(\bar\phi) - i\lambda^a\psi^i_+\partial_i J_a(\phi) - i\bar\lambda^a\bar\psi^i_+\partial_i\bar{J}_a(\bar\phi) \right).
\end{equation}
All couplings come from the E and J-type superpotentials. Note that $\phi$ is dimensionless in 2d, (and fermions are of dimension $1/2$), therefore both $E^a$ and $J_a$ should have dimension $1$. We will include an explicit coupling $\mu$ of dimension $1$ in the theory, and replace $E^a \to \mu E^a$ and $J_a \to \mu J_a$ in the above action, thinking of $E^a(\phi)$ and $J_a(\phi)$ as dimensionless functions of dimensionless fields $\phi^i$ now.

In Section \ref{OPEcoh} we saw that in order to compute the OPE of the cohomology classes we can turn off all dimensionful couplings in the theory. In particular, we can tune $\mu$ to zero. This will remove all interactions from the above action. Thus to compute the OPE of the cohomology classes, it is enough to consider the free theory:
\begin{align}
S_0 =\frac{1}{\pi}\int\d^2x \Big(-\partial_{--}\bar\phi^i\partial_{++}\phi^i -\frac{i}{2}\bar\psi^i_+\partial_{--}\psi^i_+ - i\bar\lambda^a\partial_{++}\lambda^a - \frac{1}{2}G^a\bar{G}^a\Big).
\end{align}
Its correlators can be conveniently combined into superfield correlators:
\begin{align}
\left\langle\bar\Phi^i(x,\theta')\Phi^j(y,\theta)\right\rangle &= \delta^{ij}\log\left( r^{--} r^{++} \right),\cr
\left\langle\bar\Lambda^a(x,\theta')\Lambda^b(y,\theta)\right\rangle &=\delta^{ab}\frac{i}{r^{--}},\cr
\end{align}
where
\begin{equation}
r^{--} = x^{--} - y^{--},\quad r^{++}=x^{++} - y^{++} + i \theta^+\bar\theta^+ + i\theta'^+\bar\theta'^+ + 2i\bar\theta'^+\theta^+.
\end{equation}
Now we want to compute the OPE of $\tilde\Y$ from \eqref{stressT} with itself. $\tilde\Y$ represented a candidate stress-energy tensor in the cohomology and was given by:
\begin{equation}
\tilde\Y = \sum_i\left[\left(1-\frac{\alpha_i}{2} \right)\partial_{--}\Phi^i\partial_{--}\bar\Phi^i - \frac{\alpha_i}{2}\Phi^i\partial_{--}^2\bar\Phi^i\right] + \sum_a\left[\frac{i}{2}(1+\t\alpha_a)\Lambda^a\partial_{--}\bar\Lambda^a - \frac{i}{2}(1-\t\alpha_a)\partial_{--}\Lambda^a\bar\Lambda^a\right].
\end{equation}
Using the OPE above, we find that:
\begin{equation}
\tilde\Y(x)\tilde\Y(y) \sim \frac{c/2}{(x^{--}-y^{--})^4} + \frac{2\tilde\Y(y)}{(x^{--}-y^{--})^2} + \frac{\partial_{--}\tilde\Y(y)}{x^{--}-y^{--}} + \{\bar{Q}_+,\dots\},
\end{equation}
where the notation $\{\bar{Q}_+,\dots\}$ for the unimportant term is slightly inaccurate: what we actually mean is that the term that we drop becomes $\bar{Q}_+$-exact after we put $\theta^+=\bar\theta^+=0$, but as a shorthand we will denote it as $\{\bar{Q}_+,\dots\}$. The central term is:
\begin{equation}
\label{central}
c= \sum_{i}(2 - 6\alpha_i + 3\alpha_i^2) + \sum_a (1-3\t\alpha_a^2).
\end{equation}
This matches the result of \cite{SW} and shows that we indeed have the stress-energy tensor in the cohomology.

Before we found that in case there is a $U(1)$ flavor symmetry, there is another $\bar{D}_+$-closed superfield $\F_{--}$, which gives rise to the left-moving $U(1)$ current in the cohomology. Recall that:
\begin{equation}
\F_{--} = -\sum_a\t q_a\Lambda^a\bar\Lambda^a -i\sum_i q_i\Phi^i\partial_{--}\bar\Phi^i.
\end{equation}
We can similarly compute its OPE:
\begin{equation}
\F_{--}(x)\F_{--}(y)\sim \frac{\sum_i q_i^2 - \sum_a\t q_a^2}{(x^{--}-y^{--})^2} + \{\bar{Q}_+,\dots\}.
\end{equation}

This current creates ambiguity, as we explained before: we can replace $\tilde\Y$ by $\tilde\Y + \lambda\partial_{--}\F_{--}$ for any $\lambda\in\Bbb R$ and get another stress-energy tensor in the cohomology. The unique one is picked by requiring that the $[\F_{--}]$ cohomology class be primary with respect to the correct stress-energy tensor, whenever it is possible to impose such a condition. Equivalently, since shifting by the current shifts the central charge, one can ask that the value of the central charge \eqref{central} be extremal with respect to the shifts $(\alpha_i, \t\alpha_a)\to (\alpha_i +\lambda q_i, \t\alpha_a + \lambda\t q_a)$. Any of these two criteria of course give the same equation:\footnote{While \eqref{central} is actually a left-moving central charge $c_L$, this is also equivalent to the $c_R$-extremization \cite{Benini:2012cz} since $c_L-c_R$ is fixed by the gravitational anomaly.}
\begin{equation}
\label{unique}
\sum_i q_i(1-\alpha_i) + \sum_a\t q_a\t\alpha_a =0.
\end{equation}
In a generic situation, this equation allows to pick a unique solution $(\alpha_i,\t\alpha_a)$ and write a correct stress-energy tensor. If the action admits $f$ independent $U(1)$ flavor symmetries described by charges $(q_i^n, \t q_a^n), n=1\dots f$, we should write the equation \eqref{unique} for each of them. Again, generically, one can expect this to give a condition to pick the unique stress-energy tensor in the cohomology.

However, non-generic situations are possible, when this equation might either not fix the stress-energy tensor completely, or might have no solutions at all. We will explore such examples for the $\N=(2,2)$ theories later: in such cases, indices $a$ and $i$ take the same set of values, we have $\t\alpha_a=\alpha_{i=a}$, and flavor symmetries (which should differ from the $\N=(2,2)$ R-symmetries) have $\t q_a=q_{i=a}$. Therefore the equation \eqref{unique} reduces to just $\sum_{i}q_i=0$, which either holds identically and therefore imposes no constraints on $\alpha_i$, or does not hold at all. In a former situation, the ambiguity of choosing the unique stress-energy tensor is not removed and is just present in the IR. In a latter situation, there is no solution to $\eqref{unique}$, which means that $[\F_{--}]$ cannot be made primary by choosing the proper stress-energy tensor. There is an unwanted central term in the $\tilde{\Y}\F_{--}$ OPE, which cannot be removed and signals that there is an obstruction for the IR compact CFT to exist. We will see in examples that there is a flat direction in the potential. This is usually interpreted as a lack of normalizable vacuum.

One can also note that if we decide to study the gauge theory obtained by gauging the flavor symmetry with charges $(q_i, \t q_a)$, then the above equations become related to anomalies. Namely, the central term in the $\F_{--}\F_{--}$ OPE becomes just the gauge anomaly (so it is the t'Hooft anomaly in the LG model context): we need $\sum_i q_i^2 - \sum_a \t q_a^2=0$ for the gauge theory to exist\cite{GGP2}. Then equation \eqref{unique} becomes the condition for the R-symmetry defined by charges $(\alpha_i,\t\alpha_a)$ to be non-anomalous \cite{SW}.

\subsubsection{Classical and quantum chiral algebra}
When we were discussing the OPE in the cohomology, we argued that, as a consequence of conformal invariance, there should be no dimensionful couplings present in the OPE. We can generalize that further to say that the chiral algebra should not depend on any dimensionful couplings at all. Any algebraic relations that involve dimensionful coupling constants would violate the combination of scale and Lorentz invariance.

One of the basic facts about theories we study is that they are free in the UV. In fact, this provides an alternative argument for why the singular part of the OPE is independent of couplings. Short-distance singularities of operators are simply governed by the free theory, even before passing to the $\bar{Q}_+$ cohomology (however, we find the argument based on Lorentz and scale invariance in the cohomology to be more transparent in our case).

Independence of chiral algebras on dimensionful couplings implies a useful property, which can be thought of as a sort of non-renormalization theorem. The exact quantum chiral algebra in our theories is ``almost determined'' by the classical chiral algebra (and might be called ``almost tree level exact'', although this name could be misleading). All we need to do to find the quantum counterpart is renormalize composite operators. Composite operators can be thought of as several fundamental fields brought into one point, and in the process we should subtract short-distance singularities. It might well happen (and will happen in concrete examples later) that even though the classical operator is in $\bar{Q}_+$ cohomology, the infinite piece you have to subtract is not annihilated by $\bar{Q}_+$. In this way, renormalization of composite operators representing classical cohomology classes can remove part of the classical cohomology. The claim is that what you obtain using this procedure is the exact answer.

To understand why this is true, we will think of an exact quantum theory as a set of local operators, which satisfy OPE relations and operator equations of motion. As we said, short-distance singularities are governed by the free theory, so singular part of the OPE does not care about interactions and operator equations of motion. Non-singular part of the OPE can be thought of as a definition of composite operators, and this is the point where we should be careful, as already noted before. The remaining thing we need to care about are operator equations of motion. 

If we stare at classical equations \eqref{LG_EOM}, we can understand that they do not have any short-distance singularities and can be made into operator equations. The question one might ask is whether they receive any corrections at the quantum level. If there were such corrections, they would be a result of interactions and would depend on the dimensionful coupling\footnote{In fact, the right hand side of \eqref{LG_EOM} is already proportional to $\mu$, so additional terms would be multiplied by higher powers of $\mu$} $\mu$. If this could change the answer for chiral algebra, it would mean that the algebra depends on a dimensionful constant $\mu$. We know that this is impossible on general grounds, so we expect that quantum corrections to operator equations of motion are not important for the chiral algebra computation.

In fact, thinking slightly more general, the situation might be even simpler. Suppose we have some renormalizable field theory, and we define it in the path integral approach. This means that we choose our favorite regularization to make path integral finite-dimensional, define the action and the measure in this regularization and add counterterms, if needed. Or, alternatively, think in terms of bare fields and couplings, without any counterterms. The standard way to derive equations of motion which hold under correlators, i.e., operator equations of motion, is through integration by parts. For renormalizable field theories defined in this way, these equations of motion hold exactly when written in terms bare fields. If we write them in terms of physical fields and counterterms, then counterterms of course contribute to equations of motion, but their role is to renormalize composite operators that appear in equations of motion. This becomes very clear in the example of the $\lambda\phi^4$ theory. The equation of motion of the $\lambda\phi^4$ with counterterms is:
\begin{equation}
(\Box + m^2)\phi = \lambda \phi^3 + \delta_m \phi + \delta_\phi \Box \phi + \delta_\lambda \phi^3,
\end{equation}
and by some simple manipulations with diagrams, one can see that these three terms on the right are precisely what one needs to define a composite operator $\lambda\phi^3$. The mass counterterm $\delta_m\phi$ removes singularity coming from the self-contraction in $\phi^3$, while the other two remove singularities coming from contractions between $\phi^3$ and one insertion of the interaction vertex $\lambda\phi^4/4$. It is quite obvious that this continues to higher orders of perturbation theory, simply because the theory is renormalizable and has only these three counterterms.

It is not completely clear how general this argument is and whether it holds for gauge theories, but it definitely works for our LG models. Moreover, it is possible to show that our models do not need any counterterms at all.

So our conclusion is that equations \eqref{LG_EOM} hold exactly once we properly define composite operators appearing there. This supports our claim that to compute quantum chiral algebra, we need to find the classical one and then check which part of it survives after the renormalization of composite operators.

All these statements are true in perturbation theory. They might not hold if non-perturbative corrections become relevant. For example, instantons might lift cohomology classes \cite{Y}, and this has to be studied separately. In our case we assume that the worldsheet and the target are topologically trivial, so non-perturbative corrections are not expected.

\subsubsection{Non-abelian global symmetries}
In addition to $U(1)$ global symmetries, the action may also have non-abelian linearly realized global symmetries that commute with SUSY. They generally are of the form $\Phi^i \to A^i_j\Phi^j, \Lambda^a \to B^a_b \Lambda^b$. The kinetic part of the action implies that $A\in U(N_\Phi)$ and $B \in U(N_\Lambda)$, where $N_\Phi$ is the number of chiral superfields $\Phi^i, i=1\dots N_\Phi$, and $N_\Lambda$ is the number of Fermi superfields $\Lambda^a, a=1\dots N_\Lambda$.

It is clear that by a unitary transformation $\Phi^i \to U^i_j\Phi^j$, $\Lambda^a \to V^a_b\Lambda^b$ one can always bring $A$ and $B$ into the diagonal form, and in such a basis they will describe just the $U(1)$ global symmetry. Therefore, in order to have something new compared to the previous discussion, we assume that the action has some $U(1)$ global symmetries and, on top of that, also has some non-abelian symmetries. Altogether, they close to a subgroup $G\subset U(N_\Phi)\times U(N_\Lambda)$. The free theory has the full $U(N_\Phi)\times U(N_\Lambda)$ symmetry, which is then broken to the subgroup $G$ by the $E$ and $J$ superpotentials. 

Embedding $G\subset U(N_\Phi)\times U(N_\Lambda)$ defines an $(N_\Phi + N_\Lambda)$-dimensional representation of $G$ on superfields of our model. This representation is reducible and can be decomposed as a direct sum of an $N_\Phi$-dimensional representation $R_\Phi$ on chiral superfields and an $N_\Lambda$-dimensional representation $R_\Lambda$ on Fermi superfields. Let the Hermitian generators of this subgroup in the representation $R_\Phi$ be called $t_\alpha,\,\alpha=1\dots |G|$, and in the representation $R_\Lambda$ -- $\tau_\alpha,\, \alpha=1\dots |G|$. The infinitesimal transformation is:
\begin{align}
\Phi^i &\to \Phi^i + i\epsilon^\alpha (t_\alpha)^i_j \Phi^j,\cr
\Lambda^a &\to \Lambda^a + i\epsilon^\alpha (\tau_\alpha)^a_b\Lambda^b.
\end{align}
The condition on $J$ and $E$ for this to be a symmetry is:
\begin{align}
(t_\alpha)^i_j\Phi^j\partial_i J_a(\Phi)+(\tau_\alpha)^b_a J_b(\Phi)&=0,\cr
(t_\alpha)^i_j\Phi^j\partial_i E^a(\Phi) - (\tau_\alpha)^a_b E^b(\Phi)&=0.
\end{align}
It is straightforward to repeat what we had done for abelian symmetries and to find the corresponding element in the $\bar{D}_+$-cohomology:
\begin{equation}
\J_\alpha = (\tau_\alpha)^a_b\Lambda^b\bar\Lambda^a + i(t_\alpha)^i_j\Phi^j\partial_{--}\bar\Phi^i.
\end{equation}
If we write $[t_\alpha,t_\beta]=i f_{\alpha\beta}^\gamma t_\gamma$, then the OPE of these currents is given by:
\begin{equation}
\J_\alpha(x)\J_\beta(y) \sim \frac{{\rm tr}(t_\alpha t_\beta) - {\rm tr}(\tau_\alpha \tau_\beta)}{(x^{--}-y^{--})^2} + \frac{f_{\alpha\beta}^\gamma \J_\gamma(y)}{x^{--}-y^{--}} + \{\bar{Q}_+,\dots\}.
\end{equation}
We have ${\rm tr}(t_\alpha t_\beta)=2x_\Phi \delta_{\alpha\beta}$ and ${\rm tr}(\tau_\alpha\tau_\beta)=2x_\Lambda\delta_{\alpha\beta}$, where $x_\Phi$ and $x_\Lambda$ are Dynkin indices of the representations $R_\Phi$ and $R_\Lambda$ respectively. Therefore, in the cohomology we find a current algebra of $G$ at the level $r=2(x_\Phi - x_\Lambda)$.

\section{$\N=(2,2)$ models} 

If in a general $\N=(0,2)$ LG model as described before we put $E^a=0$, take $a$ to be the same sort of index as $i$, i.e., just put $N_\Lambda=N_\Phi$ (recall that everything is topologically trivial in our discussion) and take $J_a(\Phi)=\frac{\partial W(\Phi)}{\partial\Phi^{i=a}}$ for some holomorphic superpotential $W(\Phi)$, we get a general $\N=(2,2)$ LG model. In such a case $(0,2)$ superfields are promoted to $(2,2)$ chiral superfields:
\begin{equation}
\vPh^i = \Phi^i + i\sqrt{2}\theta^-\Lambda^i -i\theta^-\bar\theta^-\partial_{--}\Phi^i.
\end{equation}

With $\N=(2,2)$ supersymmetry, we can go further in the discussion of general properties of the chiral algebra in the $\bar{Q}_+$-cohomology. First of all, let us get rid of the trivially reducible case. Suppose that we can organize superfields $\vPh^i$ into two nonempty sets: $\{\vPh^1, \vPh^2,\dots, \vPh^{s}\}$, $\{\vPh^{s+1}, \vPh^{s+2},\dots, \vPh^{N_\Phi}\}$, so that the superpotential can be written as a sum:
\begin{equation}
\label{reducible}
W(\vPh) = W^{(1)}(\vPh^1,\vPh^2,\dots,\vPh^s) + W^{(2)}(\vPh^{s+1},\vPh^{s+2},\dots,\vPh^{N_\Phi}),
\end{equation}
This superpotential just describes 2 separate LG models which do not interact with each other. The space of observables in such a model is just the graded-symmetric tensor product of the spaces for each of the two models, and the supercharge is the sum $\bar{Q}_+ = \bar{Q}^{(1)}_+ + \bar{Q}^{(2)}_+$, where each term in the sum acts on the corresponding factor in the graded-symmetric tensor product. It is a simple algebraic exercise to prove that the cohomology of such a $\bar{Q}_+$ is just the graded-symmetric tensor product of the cohomologies of $\bar{Q}_+^{(1)}$ and $\bar{Q}^{(2)}_+$.

Therefore, without any loss of generality, it is enough to study superpotentials which cannot be decomposed as in \eqref{reducible}, and can never be brought into such a decomposable form by a holomorphic change of coordinates on the target. We will assume this from now on. Note that it was shown in \cite{Mel2} that with such an assumption, no accidents happen in the IR, which also simplifies life a lot.

Since we put $E^a=0$, quasihomogeneity conditions \eqref{quasih} now always have at least one solution, $\tilde\alpha_a=1, \forall a$, $\alpha_i=0, \forall i$. Therefore, according to our previous discussion, there is always a stress-energy tensor in the cohomology. It is interesting, however, to study the case when $W(\vPh)$ is quasi-homogeneous itself:
\begin{equation}
\sum_i \beta_i\Phi^i \frac{\partial W}{\partial\Phi^i} = W(\Phi).
\end{equation}
After all, as was noted in \cite{VW}, this is the case most relevant for studying the IR fixed point of the LG model. With this property, if we take $\alpha_i=\beta_i$ and $\tilde\alpha_a = \beta_{i=a}$, we get another solution of \eqref{quasih}. In other words, there exists a $U(1)$ flavor symmetry corresponding to the solution $q_i=\beta_i$, $\tilde q_a=\beta_{i=a}-1$ of \eqref{quasih2}.

If there is only one such flavor symmetry, we can see that the equation \eqref{unique} picks $\alpha_i=\beta_i$ and $\tilde\alpha_a = \beta_{i=a}$ as defining the correct stress-energy tensor. Indeed, these values satisfy \eqref{unique}, while another solution, $\tilde\alpha_a=1,\,\alpha_i=0$, inserted in \eqref{unique}, gives $\sum_i q_i + \sum_a\tilde q_a = \sum_i (2\beta_i -1)$, which is generically non-zero. The last sum being zero corresponds to various degenerate cases, for example if superpotential is just a quadratic polynomial (which means that all fields are massive, the IR theory is trivial and the chiral algebra should be trivial too). We will not concentrate on such cases.

On the other hand, there can be more flavor symmetries in the model:
\begin{equation}
\vPh^i \to e^{-i\gamma_i\epsilon}\vPh^i,
\end{equation}
if one can find such a system of charges $\gamma_i$ that:
\begin{equation}
\label{flavor22}
\sum_i \gamma_i \Phi^i \frac{\partial W}{\partial\Phi^i}=0.
\end{equation}
This gives a solution $q_i=\gamma_i$, $\tilde q_a=\gamma_{i=a}$ of \eqref{quasih2}. Note that both the solution $q_i=\beta_i$, $\tilde q_a=\beta_{i=a}-1$ and the solution $q_i=\gamma_i$, $\tilde q_a=\gamma_{i=a}$ describe flavor symmetries from the $\N=(0,2)$ point of view, since they just satisfy \eqref{quasih2}. However, from the $\N=(2,2)$ point of view, only the latter one is a flavor symmetry, while the former one becomes the left-handed R-symmetry of the $\N=(2,2)$ SUSY, which is seen from the fact that $\Phi$'s and $\Lambda$'s charges differ by one.

The action of the LG model in the $(2,2)$ superspace is:
\begin{equation}
S=\frac{1}{4\pi}\int\d^2x \d^4\theta\, \bar\vPh^i\vPh^i + \frac{1}{4\pi}\int \d^2x\d^2\theta\, W(\vPh) + \frac{1}{4\pi}\int\d^2x\d^2\bar\theta\, \bar{W}(\bar\vPh)
\end{equation}
The superspace equations of motion are simply:
\begin{equation}
\bar{D}_+\bar{D}_-\bar\vPh^i = \frac{\partial W}{\partial \vPh^i}.
\end{equation}
As was first noted in \cite{W}, we can find an element in the $\bar{D}_+$-cohomology represented by the $(2,2)$ superfield:
\begin{equation}
\J = \sum_i \left( \frac{1-\beta_i}{2}D_-\vPh^i\bar{D}_-\bar\vPh^i - i\beta_i\vPh^i \partial_{--}\bar\vPh^i \right),
\end{equation}
which can then be expanded in components with respect to $\theta^-$ and $\bar\theta^-$: the lowest component is the left-handed R-current, the top component is the stress-energy tensor (which, using our earlier $\N=(0,2)$ terminology, corresponds to the solution $\alpha_i=\beta_i$, $\tilde\alpha_a=\beta_{i=a}$ of \eqref{quasih}), and the fermionic components are the two left-handed supersymmetries. Therefore, this $\J$ generates a left-moving $\N=2$ superconformal algebra in the $\bar{D}_+$-cohomology.

If there exist additional $U(1)$ flavor symmetries characterized by weights $\gamma_i$ satisfying \eqref{flavor22}, then there is another $\bar{D}_+$-cohomology class represented by:
\begin{equation}
\label{fl22}
\Psi = \frac{1}{2}\sum_i \gamma_i\vPh^i\bar{D}_- \bar\vPh^i,
\end{equation}
so the derivative:
\begin{equation}
D_-\Psi = \sum_i \gamma_i\left( \frac{1}{2}D_-\vPh^i\bar{D}_-\bar\vPh^i + i\vPh^i\partial_{--}\bar\vPh^i \right)
\end{equation}
generates ambiguity, because we can replace $\J \to \J + \lambda D_-\Psi,\, \forall \lambda\in\Bbb{R}$. Of course, this is still the same ambiguity of the $\N=(0,2)$ stress-tensor multiplet related to $U(1)$ flavor symmetries that we were discussing before. The only difference is that by now we have dealt with the $U(1)$ global symmetry which is the left-handed R-symmetry from the $\N=(2,2)$ point of view (it was described by the charges $q_i=\beta_i$, $\tilde q_a=\beta_{i=a}-1$), and what we are left with in \eqref{fl22} corresponds to the actual $\N=(2,2)$ flavor symmetry. Similar to what we had for a more general $\N=(0,2)$ case, we could have analyzed this ambiguity using the $\N=(2,2)$ supersurrent multiplet, especially since its structure is described in details in the Appendix C of \cite{DS}. However, we chose not to do this, as it would not give us anything essentially new compared to what we have already understood.

Previous discussion of the OPE in the cohomology being determined by the free propagators of course still holds. The free propagator of chiral superfields is:
\begin{equation}
\left\langle \bar\vPh^i(x_1,\theta_1) \vPh^j(x_2,\theta_2) \right\rangle = \delta^{ij}\log\left( R^{--}_{12} R^{++}_{12} \right),
\end{equation}
where
\begin{align}
R_{12}^{--} &= x_1^{--} - x_2^{--}+ i\theta_1^-\bar\theta_1^- + i\theta_2^-\bar\theta_2^- + 2i\bar\theta_1^-\theta_2^-,\cr R_{12}^{++}&=x_1^{++} - x_2^{++} + i \theta_1^+\bar\theta_1^+ + i\theta_2^+\bar\theta_2^+ + 2i\bar\theta_1^+\theta_2^+.
\end{align}
We can compute the OPEs:
\begin{align}
\label{n2alg}
\J(x_1,\theta_1)\J(x_2,\theta_2) \sim -\frac{c}{3(\mathbf{r}_{12})^2} - \frac{2\theta^-_{12}\bar\theta^-_{12}}{(\mathbf{r}_{12})^2}\J(x_2,\theta_2) - \frac{i\theta^-_{12}}{\mathbf{r}_{12}}D_-\J(x_2,\theta_2)\cr
- \frac{i\bar\theta^-_{12}}{\mathbf{r}_{12}}\bar{D}_-\J(x_2,\theta_2)
-\frac{2\theta^-_{12}\bar\theta^-_{12}}{\mathbf{r}_{12}}\partial_{--}\J(x_2,\theta_2) + \{\bar{Q}_+,\dots\},
\end{align}
where
\begin{align}
\theta^-_{12}&=\theta^-_1 - \theta^-_2,\quad \bar\theta^-_{12}=\bar\theta^-_1 - \bar\theta^-_2,\cr
\mathbf{r}_{12} &= x_1^{--} - x_2^{--} + i\bar\theta^-_1\theta^-_2 - i\bar\theta^-_2\theta^-_1,\cr
\end{align}
and the central charge is:
\begin{equation}
\label{central22}
c = 3\sum_i (1-2\beta_i).
\end{equation}
Equation \eqref{n2alg} encodes the $\N=2$ superconformal algebra with the central charge $c$ (this equation, but in slightly different conventions, was present in \cite{W}). Of course we could have obtained the same value of the central charge using the more general equation \eqref{central}, which holds for more general $\N=(0,2)$ LG models. One would have to put $\alpha_i=\beta_i$, $\tilde\alpha_a=\beta_{i=a}$ there.

Notice that the central charge \eqref{central22} is linear in $\beta_i$. This means that if we have $U(1)$ flavor symmetries such that \eqref{flavor22} holds, we can no longer get rid of the ambiguity $\beta_i \to \beta_i + \lambda\gamma_i$ by simply asking the central charge to take the extremal value. This is related to the fact that the OPE of the cohomology class represented by \eqref{fl22} with itself is regular:
\begin{equation}
\Psi(x_1,\theta_1)\Psi(x_2,\theta_2) \sim \{\bar{Q}_+,\dots\}.
\end{equation}
So that the OPE of $[\Psi]$ with $[\J]$ is the same as with $[\J + \lambda D_-\Psi],\, \forall\lambda\in\Bbb{R}$. The $\J\Psi$ OPE is:
\begin{align}
\label{JFope}
\J(x_1,\theta_1)\Psi(x_2,\theta_2) \sim \kappa\frac{\theta^-_{12}}{(\mathbf{r}_{12})^2} - \frac{\theta^-_{12}\bar\theta^-_{12}}{(\mathbf{r}_{12})^2}\Psi(x_2,\theta_2) - \frac{i\theta^-_{12}}{\mathbf{r}_{12}}D_-\Psi(x_2,\theta_2)\cr
-\frac{2\theta^-_{12}\bar\theta^-_{12}}{\mathbf{r}_{12}}\partial_{--}\Psi(x_2,\theta_2)-\frac{i}{\mathbf{r}_{12}}\Psi(x_2,\theta_2) + \{\bar{Q}_+,\dots\},
\end{align}
where $\kappa=\sum_i\gamma_i$. Compare this with what one expects for the OPE of $\J$ with some superconformal primary superfield $\P$:
\begin{align}
\label{primary}
\J(x_1,\theta_1)\P(x_2,\theta_2) \sim  - \frac{2\theta^-_{12}\bar\theta^-_{12}}{(\mathbf{r}_{12})^2}\Delta\P(x_2,\theta_2) - \frac{i\theta^-_{12}}{\mathbf{r}_{12}}D_-\P(x_2,\theta_2)\cr
- \frac{i\bar\theta^-_{12}}{\mathbf{r}_{12}}\bar{D}_-\P(x_2,\theta_2) -\frac{2\theta^-_{12}\bar\theta^-_{12}}{\mathbf{r}_{12}}\partial_{--}\P(x_2,\theta_2)-\frac{i}{\mathbf{r}_{12}}q \P(x_2,\theta_2) + \{\bar{Q}_+,\dots\},
\end{align}
where $\Delta$ is the conformal dimension of $\P$ and $q$ is its R-charge.
What we see is that for non-zero values of $\kappa$, $\Psi$ is non-primary, and moreover it is not a descendant of any primary, as can be seen from unitarity and global superconformal invariance of the vacuum of the IR theory.\footnote{\label{fnoteJP}The argument is as follows. Presence of the central term in \eqref{JFope} implies through the operator-state correspondence that there is a state $|\psi\rangle$ in the IR CFT such that $G_{+1/2}^-|\psi\rangle=\kappa|0\rangle + \bar{Q}_+|\phi\rangle$, where $G^-_{+1/2}$ is one of the superconformal generators, $|0\rangle$ is the vacuum state and $|\phi\rangle$ is some state. Taking the dimension-zero component of this equality, we can assume that $\psi$ has dimensions $(1/2,0)$, so that $G_{+1/2}^-|\psi\rangle$ has dimension zero. Since in a unitary theory there are no operators of negative dimension, $\bar{Q}_+|\phi\rangle$ should not be there: $G_{+1/2}^-|\psi\rangle=\kappa|0\rangle$. Invariance of the vacuum implies $G^+_{-1/2}G^-_{+1/2}|\psi\rangle=0$. Since in a unitary theory $(G^-_{+1/2})^\dagger=G^+_{-1/2}$, by multiplying with $\langle\psi|$, the last equality implies $G^-_{+1/2}|\psi\rangle=0$, which gives a contradiction unless $\kappa=0$. } Therefore, the non-zero $\kappa$ becomes an obstruction for the IR CFT to exist. This happens for example in a model with two superfields $X$ and $Y$ and superpotential $W=XY^2$.

Notice that since the $\Psi\Psi$ OPE is regular, so is the $D_-\Psi D_-\Psi$ OPE. The absence of central term in it means, as we have mentioned in Section \ref{OPcoh}, that the corresponding flavor symmetry can be gauged without encountering gauge anomalies. Possible non-zero value of $\kappa$ then becomes the anomaly for the right-handed R-symmetry. This would be relevant if we were studying gauge theories.

The theory can also have non-abelian flavor symmetries, which lead, as we have argued before, to the current algebra in the cohomology. In our discussion of general $(0,2)$ theories, the level of this current algebra was given by the difference of Dynkin indices: $r=2x_\Phi - 2x_\Lambda$. The first term here corresponded to the way flavor symmetry acted on $\Phi$'s, and the second -- on $\Lambda$'s. In the $(2,2)$-supersymmetric case, the flavor symmetry acts in the same way on $\Phi$'s and $\Lambda$'s, as they are just components of the $(2,2)$ chiral superfields $\vPh^i$. So $x_\Phi = x_\Lambda$. We conclude that the current algebra in the cohomology corresponding to some flavor symmetry of the $\N=(2,2)$ supersymmetric LG model always has level zero.

\section{Examples}
In this section we will consider a few examples of applications of our machinery to the $\N=(2,2)$ LG models, where we can say something about the chiral algebra and therefore draw some conclusions about the theory to which the model flows in the IR.
\subsection{Degenerate examples}\label{degen}
Consider the theory of two chiral superfields $X$ and $Y$ with superpotential
\begin{equation}
W = XY^2.
\end{equation}
This theory has a non-trivial flavor symmetry. A possible charge assignment is: $\gamma_X=2$, $\gamma_Y=-1$, so that
\begin{equation}
\gamma_X X \frac{\partial W}{\partial X} + \gamma_Y Y \frac{\partial W}{\partial Y}=0.
\end{equation}
As we know from the equation \eqref{fl22}, there is an extra operator $\Psi$ in the cohomology as a result of this flavor symmetry. Since $\gamma_X + \gamma_Y=1\ne 0$, the OPE \eqref{JFope} tells us that this operator is not primary. Moreover, as we explained in the Footnote \ref{fnoteJP}, an operator satisfying \eqref{JFope} cannot be made primary in a unitary CFT with invariant vacuum. Therefore, its existence indicates that the RG flow does not end at any compact CFT: the deep IR theory does not have a normalizable vacuum. In fact, the superpotential has a flat direction $Y=0$, and we can conclude that the low-energy mode describing propagation along this flat direction renders vacuum non-normalizable. One can get a compact conformal fixed point if we add a perturbation $\epsilon X^{2n+1}$ to the superpotential. This actually corresponds to having the $D$ series of minimal models at the IR fixed point, with the exact choice of the model depending on $n$, even for small $\epsilon$. By sending $\epsilon \to  0$, the compact IR fixed point will most likely go to infinity, signaling that the degenerate theory $W=XY^2$ will behave differently. It would be great to have a better understanding of such non-compact IR theories.

This flavor symmetry could be gauged, however, as we have noted before, this would make right-handed R-symmetry anomalous because of $\gamma_X + \gamma_Y\ne 0$.

By considering a slightly different superpotential, namely:
\begin{equation}
W = X^2 Y^2,
\end{equation}
we get again a theory with flavor symmetry, but the charges now can be chosen as $\gamma_X=1$, $\gamma_Y=-1$, so that $\gamma_X + \gamma_Y=0$. Therefore, the bad central term does not appear in \eqref{JFope}, and the theory should have a better IR limit, even though the superpotential still has flat directions. For example, if we gauge this flavor symmetry, the flat directions are removed and we still get a theory with the right-handed R-symmetry. We are not going to study this example any further.

\subsection{$\N=2$ minimal models}
A series of $\N=(2,2)$ LG models are known to flow in the IR to the $\N=(2,2)$ minimal models. These superconformal theories are relatively simple. The central charge is given by \cite{N2min1, N2min1.a, N2min1.b, N2min1.c, N2min1.5}:
\begin{equation}
c=\frac{3k}{k+2},\, k\ge 1,
\end{equation}
and there is a known spectrum of possible superconformal primaries. The A-D-E classification of modular-invariant theories is known \cite{N2min2, N2min2.a, N2min3, N2min3.a}, and the corresponding LG superpotentials have been identified before. So, we can try to compute the chiral algebra both for the LG model and for the minimal model which is supposed to arise in the IR, therefore providing more evidence for this relation, as well as demonstrating the power of chiral algebras.

\subsubsection{The $A_{k+1}$ series}
For a given $k$, the diagonal $A_{k+1}$ minimal model is the simplest one. Its set of primaries has a subset of $k+1$ fields which are chiral primary with respect to ${\rm \tilde{SVir}}$. Let us call them $\O_s,\, s=0,\dots k$, where $\O_0=1$ is the identity operator and $\O_s$ has left-right conformal dimensions $(h,\bar{h})=(\frac{s}{2(k+2)},\frac{s}{2(k+2)})$ and left-right $U(1)$ charges $(q,\bar{q})=(\frac{s}{k+2},\frac{s}{k+2})$. As we see, they all are chiral primaries with respect to both ${\rm SVir}$ and ${\rm \tilde{SVir}}$. Therefore, together with the $\N=2$ currents, they generate the chiral algebra of the theory, as well as the anti-chiral algebra abtained analogously by taking the cohomology of $\bar{Q}_-$. 

We expect to get the same result from the LG model description. It is obtained by considering only one chiral superfield $\vPh$ with the superpotential:
\begin{equation}
W(\vPh) = \frac{\vPh^{k+2}}{k+2}.
\end{equation}
The equations of motion are 
\begin{align}
\bar{D}_+\bar{D}_-\bar\vPh - \vPh^{k+1}&=0,\cr
D_- D_+ \vPh - \bar\vPh^{k+1}&=0.
\end{align}
Differentiating these equations and multiplying them by arbitrary polynomials of $\vPh$, $\bar\vPh$ and their derivatives, we get a differential ideal $\I$.
The algebra $\hat\F_0$ consists of arbitrary polynomials of variables $\partial_+^n \partial_-^m D_+^k D_-^p\vPh$ and $\partial_+^n \partial_-^m \bar{D}_+^k \bar{D}_-^p\bar\vPh$ for non-negative integers $n, m, k, p$, modulo the ideal $\I$:
\begin{equation}
\hat\F_0 = \C[\dots, \partial_+^n \partial_-^m D_+^k D_-^p\vPh, \partial_+^n \partial_-^m \bar{D}_+^k \bar{D}_-^p\bar\vPh,\dots]/\I.
\end{equation} 
It is not hard to find another set of generators, which will generate $\hat\F_0$ as a super-commutative  polynomial algebra itself. We already explained it in the context of general $\N=(0,2)$ LG models. Namely, we can take:
\begin{equation}
\mathcal{G}=\{\partial_{--}^n\vPh, \partial_{++}^n\vPh, D_-\partial_{--}^n\vPh, D_+\partial_{++}^n\vPh,\partial_{--}^n\bar\vPh, \partial_{++}^n\bar\vPh, \bar{D}_-\partial_{--}^n\bar\vPh, \bar{D}_+\partial_{++}^n\bar\vPh, n\ge 0 \}.
\end{equation}
All other derivatives of elementary superfields $\vPh$ and $\bar\vPh$ can be expressed, using equations of motion, as polynomials of these generators, and moreover, there are no further algebraic relations between these generators. So we have:
\begin{equation}
\hat\F_0 \simeq \C[\mathcal{G}].
\end{equation}
We will first compute the classical cohomology of $\bar{D}_+$ acting on this space. After that we will check which part of it survives at the quantum level, when we take care to subtract singular parts from composite operators. It is clear that the cohomology classes can only be destroyed by this subtraction. Indeed, suppose we define:
\begin{equation}
:AB:(z) =\lim_{\epsilon\to 0}\left( A(z+\epsilon)B(z) - ({\rm singular\, in\, }\epsilon)\right).
\end{equation}
If $AB$ was classically in the cohomology but the singular part is not $\bar{D}_+$-closed, the operator $:AB:$ is no longer in the cohomology. If $AB$ was not in the cohomology even classically, then neither is $:AB:$, which is quite obvious. Finally, if $AB$ was classically $\bar{D}_+$-exact, then there is no need to consider $:AB:$. Even if the singular part represented some non-trivial quantum cohomology class, we would find it by starting with some other classical cohomology class anyways. So, we will look for the classical cohomology first, and then check which part of it survives subtraction of singularities.\footnote{In fact, computation of the classical cohomology is a hard combinatorial problem, while we are really only interested in quantum cohomology. So we will not determine the classical cohomology completely, only partly. As we will see, there is an $\N=2$ super-Virasoro algebra in cohomology, so our approach will be to look for those classical cohomology classes which have a chance to be superconformal primaries at the quantum level.}

To find how $\bar{D}_+$ acts on $\hat\F_0$ in terms of the generators, we act with $\bar{D}_+$ on the generators from the set $\mathcal{G}$ and, using the equations of motion, express the result in terms of these generators again. To explicitly describe $\bar{D}_+$, it is convenient to write it as a sum:
\begin{equation}
\bar{D}_+ = d_0 + d_1,
\end{equation}
where $d_0$ acts as follows:
\begin{align}
\label{d0}
d_0: &\partial_{--}^n\vPh \mapsto 0, \quad\partial_{++}^n\vPh \mapsto 0,\quad  D_-\partial_{--}^n\vPh\mapsto 0, \quad \partial_{--}^{n+1}\bar\vPh \mapsto 0, \quad\bar{D}_+\partial_{++}^n\bar\vPh\mapsto 0,\cr
& \bar{D}_-\partial_{--}^n\bar\vPh\mapsto 0,\quad D_+\partial_{++}^n\vPh \mapsto 2i\partial_{++}^{n+1}\vPh, \quad \partial_{++}^n\bar\vPh \mapsto \bar{D}_+\partial_{++}^n\bar\vPh,
\end{align}
and $d_1$ acts as:
\begin{align}
\label{d1}
d_1: &\partial_{--}^n\vPh \mapsto 0, \quad\partial_{++}^n\vPh \mapsto 0,\quad  D_-\partial_{--}^n\vPh\mapsto 0, \quad D_+\partial_{++}^n\vPh \mapsto 0, \quad \partial_{++}^n\bar\vPh \mapsto 0,\cr
&\bar{D}_+\partial_{++}^n\bar\vPh\mapsto 0,\quad \bar{D}_-\partial_{--}^n\bar\vPh\mapsto \partial_{--}^n(\vPh^{k+1}), \quad \partial_{--}^{n+1}\bar\vPh \mapsto \frac{i}{2} D_-\partial_{--}^n(\vPh^{k+1}).
\end{align}
This explicitly describes how $\bar{D}_+$ acts on the generators, and then extends to the full algebra $\hat\F_0$ by linearity and Leibniz rule. Notice that $d_0$ is just the $\bar{D}_+$ in the theory with zero superpotential, while $d_1$ includes corrections due to the superpotential. This splitting of $\bar{D}_+$ is motivated by a perturbative computation of the $\bar{D}_+$-cohommology, i.e., the spectral sequence, which we are about to perform.

Let us introduce a filtration degree on $\hat\F_0$ by saying that for generators:
\begin{equation}
\forall x\in \mathcal{G},\, {\rm fdeg}(x) = 1,
\end{equation}
which then extends multiplicatively on the whole $\hat\F_0$. We then define:
\begin{equation}
\hat\F_0^{(p)} = \{S\in\hat\F_0: {\rm fdeg}(S)\ge p\},
\end{equation}
which gives a filtration:
\begin{equation}
\hat\F_0 \cong \hat\F_0^{(0)} \supset \hat\F_0^{(1)} \supset \hat\F_0^{(2)} \supset \dots
\end{equation}
Our differential $\bar{D}_+$ obviously preserves this filtration. In particular, $d_0$ does not change the filtration degree, while $d_1$ increases it by $k$, if $k>0$. This allows us to apply spectral sequences to compute the cohomology of $\bar{D}_+$. But before that we will mention a trivial technical lemma we will need later.

\textbf{Lemma 5.1: } Let $V$ be a $\Bbb{Z}_2$-graded vector space and $S(V) = \oplus_{k\ge 0} S^k(V)$ be the graded-symmetric algebra of $V$. If there is a degree-1 differential $d: V\to V$, i.e., $d^2=0$, then by the Leibniz rule it extends to a differential acting on the graded-symmetric algebra $d: S(V) \to S(V)$, and moreover, its cohomology is:
\begin{equation}
H(S(V), d) = S \left(H(V,d)\right).
\end{equation}

Now, having this Lemma, we will proceed to compute the cohomology of $\bar{D}_+$. 

First let us consider the trivial case $k=0$. Then both $d_0$ and $d_1$ do not change the filtration degree. We can define a vector space spanned by the elements of $\mathcal{G}$: $V={\rm Span}(\mathcal{G})$. Since $\bar{D}_+=d_0+d_1$ does not change the filtration degree, it acts as a linear operator on this $V$. Next we notice that $\hat\F_0 \simeq S(V)$, so by the Lemma $H(\hat\F_0, \bar{D}_+)= S\left( H(V, \bar{D}_+) \right)$. To compute the cohomology of $\bar{D}_+$ acting as a linear operator on $V$, we notice that all elements of $\mathcal{G}$ are either not $\bar{D}_+$-closed or are $\bar{D}_+$-exact as a consequence of the equation of motion $\bar{D}_+\bar{D}_-\bar\vPh=\vPh$. So the cohomology is trivial for $k=0$ (stress-energy supercurrent $\J$ becomes $\bar{D}_+$-exact for $k=0$ ad well). 
This could be expected because the $k=0$ model is massive, and therefore the IR theory it flows to is empty.

Now, suppose $k>0$. Then at the zeroth order of spectral sequence we have:
\begin{equation}
E_0^p = \hat\F_0^{(p)}/\hat\F_0^{(p+1)},\quad E_0\equiv {\rm Gr}(\hat\F_0) \cong \bigoplus_{p\ge 0} E_0^p,
\end{equation}
where ${\rm Gr}(\hat\F_0)$ is the graded space associated with the filtered space $\hat\F_0$, and the differential acting on it is just $d_0$, which preserves grading. We note that $E_0^1\simeq {\rm Span}(\mathcal{G})$, the vector space spanned by the generators from $\mathcal{G}$ (which all have degree 1). Since $d_0$ preserves grading and, as one can easily see, $E_0 \simeq S(E_0^1)$, we just apply Lemma and get $H(E_0, d_0)= S\left(H(E_0^1, d_0)\right)$. By inspecting equations \eqref{d0}, we easily find the cohomology of $d_0$ acting on $E_0^1$. The answer is $H(E_0^1, d_0)={\rm Span}(S_0)$, where the set $S_0$ is:
\begin{equation}
S_0 = \{\partial_{--}^n\vPh,\, D_-\partial_{--}^n\vPh,\, \partial_{--}^{n+1}\bar\vPh,\, \bar{D}_-\partial_{--}^n\bar\vPh,\, n\ge 0\}.
\end{equation}
Therefore, we find the first term of the spectral sequence:
\begin{equation}
E_1 = H(E_0, d_0)\simeq \C[S_0].
\end{equation}

Now, if $k=1$, then for the first step of spectral sequence, $d_1$ becomes the differential acting on $E_1$. If $k>1$, then the differential acting on $E_1$ is just zero, and $E_2=H(E_1, 0)\simeq E_1$. Next, if $k>2$, we find that $E_3\simeq E_1$, and so on. This procedure goes on until we get to the $k$-th term of spectral sequence: $E_k\simeq E_1$. As we know from spectral sequences, the differential acting on $E_k$ should be the degree-$k$ part of $\bar{D}_+$, i.e., $d_1$. So for the next term we have:
\begin{equation}
E_{k+1} \simeq H(E_1, d_1).
\end{equation}
Since there are no components of $\bar{D}_+$ of degree higher than $k$, the spectral sequence collapses here and we conclude that:
\begin{equation}
H(\hat\F_0, \bar{D}_+) \simeq H(E_1, d_1) \simeq H(\C[S_0], d_1).
\end{equation}
So all we need to do now is compute the cohomology of $d_1$ acting on $\C[S_0]$. The way $d_1$ acts on the elements of $S_0$ is:
\begin{align}
d_1:\quad  & \partial_{--}^n\vPh \mapsto 0,\quad D_-\partial_{--}^n\vPh\mapsto 0,\cr
&\partial_{--}^{n+1}\bar\vPh \mapsto \frac{i}{2} D_-\partial_{--}^n (\vPh^{k+1}),\quad \bar{D}_-\partial_{--}^n\bar\vPh \mapsto \partial_{--}^n(\vPh^{k+1}).
\end{align}
Even though we have considerably simplified the original problem, the direct computation of the $d_1$ cohomology is still too nasty. We can simplify it more by recalling that we already have a stress-energy supercurrent in the cohomology, and therefore it is enough to look for its superconformal primaries only. Our superpotential is of a quasi-homogeneous class, with $\beta=\frac{1}{k+2}$, so the stress-energy supercurrent is:
\begin{equation}
\J = \frac{k+1}{2(k+2)}D_-\vPh \bar{D}_-\bar\vPh -\frac{i}{k+2}\vPh\partial_{--}\bar\vPh
\end{equation}
and the corresponding central charge is $c=\frac{3k}{k+2}$. Now suppose we found some polynomial $P\in \C[S_0]$ which represents a $\bar{D}_+$-cohomology class. We have the following technical Lemma:

\textbf{Lemma 5.2: } Every $d_1$-cohomology class $[P]$ which is a superconformal primary with respect to $\J$, can be represented as a polynomial of $\vPh$, $D_-\vPh$, $\bar{D}_-\bar\vPh$ and $\partial_{--}\bar\vPh$, that is $P\in\C[\vPh, D_-\vPh, \bar{D}_-\bar\vPh,\partial_{--}\bar\vPh]$.

The idea is that having higher derivatives of $\vPh$ and $\bar\vPh$ in the expression for $P$ will result in higher poles in the $\J(x,\theta) P(0,0)$ OPE, which should not be there if $[P]$ is primary. Elegant proof of this statement is not available at the moment, but calculations seem to show that it is true, so we leave it as a conjecture.

The operators $\vPh$, $\vPh^2,\dots$ $\vPh^k$ are all in the cohomology and are primaries -- we will write their OPE's with $\J$ later. $\vPh^{k+1}$ is exact and so is not in the cohomology, so any polynomial of $\vPh$ is just a linear combination of $1, \vPh, \vPh^2, \dots, \vPh^k$ in the cohomology. Since $D_-(P(\vPh))=P'(\vPh)D_-\vPh$ and $(D_-\vPh)^2=0$, any polynomial of $\vPh$ and $D_-\vPh$ is $A(\vPh) + D_- B(\vPh)$, where the second term is a descendant. Let us figure out now if there are any other primaries in the cohomology. We try to construct $d_1$-closed (or equivalently, $\bar{D}_+$-closed) polynomials from $\vPh, D_-\vPh, \bar{D}_-\bar\vPh$ and $\partial_{--}\bar\vPh$, which are not just polynomials of $\vPh$ and $D_-\vPh$. A simple computation shows that the most general such combination with even statistics is:
\begin{equation}
\mathcal{E} = \sum_{n=0}^\infty P_n(\vPh)(\partial_{--}\bar\vPh)^n\left[\frac{(n+1)(k+1)}{2}D_-\vPh\bar{D}_-\bar\vPh - i\vPh\partial_{--}\bar\vPh \right],
\end{equation}
where $P_n$ are arbitrary polynomials, while the most general odd closed element is:
\begin{equation}
\mathcal{O} = \sum_{n=1}^\infty C_n(\vPh)(\partial_{--}\bar\vPh)^nD_-\vPh,
\end{equation}
where again $C_n$ are arbitrary polynomials. 

To slightly simplify computations, we notice that since the operator $d_1$ increases the introduced above filtration degree ${\rm fdeg}$ by $k$, one can grade the cohomology by this degree, and it is enough to assume that $\mathcal{E}$ has a given fixed degree (i.e., it is a homogeneous polynomial). Next, we notice that we could introduce another grading -- by the number of derivatives in the expression. If we assign the bosonic derivative $\partial_{--}$ a ``derivative degree'' $1$ and the fermionic derivatives $D_-$ and $\bar{D}_-$ a ``derivative degree'' $1/2$, we can see that the operator $d_1$ actually lowers the ``derivative degree'' by $1/2$. Therefore, again, we can grade the cohomology by this degree, and it is enough to study the cohomology within the sector with a given ``derivative degree''. Fixing values of these two degrees -- the filtration degree and the ``derivative degree'' -- we see that it is enough, without loss of generality, to consider:
\begin{align}
\mathcal{E}^{s,n} &= \vPh^s (\partial_{--}\bar\vPh)^n\left[\frac{(n+1)(k+1)}{2}D_-\vPh\bar{D}_-\bar\vPh - i\vPh\partial_{--}\bar\vPh \right],\cr
\mathcal{O}^{s,n} &= \vPh^s (\partial_{--}\bar\vPh)^n D_-\vPh.
\end{align}
where $s$ and $n$ are non-negative integers. A simple calculation gives:
\begin{equation}
D_-\mathcal{E}^{s,n}=-i\left[ s+1 + (n+1)(k+1) \right]\mathcal{O}^{s,n+1}.
\end{equation}
This suggests that any odd element of the above form $\mathcal{O}^{s,n+1}$ that we could have possibly found in the cohomology would always by a descendant of some even element. This is also true for $\mathcal{O}^{s,0}=\vPh^s D_-\vPh = \frac{1}{s+1} D_- \vPh^{s+1}$. Therefore, it is enough to study the expression $\mathcal{E}^{s,n}$ given above. Can it represent a nontrivial cohomology class, and can it be a superconformal primary?

\paragraph{Observables $\mathcal{E}^{s,n}$ and their lifting\\}

Notice that for $s\ge k$:
\begin{equation}
d_1\left[ \vPh^{s-k}(\partial_{--}\bar\vPh)^{n+1}\bar{D}_-\bar\vPh \right] = i \mathcal{E}^{s,n},
\end{equation} 
so $\mathcal{E}^{s,n}$ is exact for $s\ge k$.
On the other hand, for $s<k$, $\mathcal{E}^{s,n}$ is obviously not exact, because, as we can see from the equation \eqref{d1}, the image of $d_1$ always contains the field $\vPh$ at least $k+1$ times, while $\mathcal{E}^{s,n}$ contains it $s+1$ times. So we conclude that $\mathcal{E}^{s,n}$ for $s<k$ indeed represents a non-trivial classical cohomology class.

Classical observables $\mathcal{E}^{s,n}$ satisfy the following multiplication rule:
\begin{equation}
\mathcal{E}^{s,n}\mathcal{E}^{t,m}=-i\mathcal{E}^{s+t+1,n+m+1}.
\end{equation}
They can be combined with the observables $\vPh^s$, for which we have:
\begin{equation}
\vPh^s \mathcal{E}^{t,n} = \mathcal{E}^{s+t,n}.
\end{equation}
We see that $\vPh$ and $\mathcal{E}^{s,n}$ generate a closed sector in the classical cohomology. As we will find soon, these are not all classical cohomology classes, there exist more. But all observables that have a chance of being superconformal primaries in the quantum cohomology are within this sector.

The stress-energy supercurrent $\J$ that we identified before is of course among these observables:
\begin{equation}
\J = \frac{1}{k+2}\mathcal{E}^{0,0}.
\end{equation}
In particular:
\begin{equation}
\J \mathcal{E}^{s,n} = -\frac{i}{k+2}\mathcal{E}^{s+1,n+1}.
\end{equation}
This equation implies that the only observables which have a chance of being superconformal primaries at the quantum level are $\mathcal{E}^{s,0}$ and $\mathcal{E}^{0,n}$. But because of:
\begin{equation}
\mathcal{E}^{s,0} = (k+2)\J\vPh^s,
\end{equation}
the former are simply descendants of $\vPh^s$. So we only have $\mathcal{E}^{0,n}$ left.

Can $\mathcal{E}^{0,n}$ represent cohomology classes in quantum theory? It turns out that only for $n=0$. The reason is that for $n>0$, the infinite piece that one has to subtract in order to define the composite operator $\mathcal{E}^{0,n}$ is not $\bar{Q}_+$-closed. 

Consider the simplest operator $\mathcal{E}^{0,1}$. We call its lowest component $e_1$:
\begin{equation}
e_1 = (k+1)\partial_{--}\bar\phi \psi_- \bar\psi_- - i \phi(\partial_{--}\bar\phi)^2.
\end{equation}
This is a composite operator whose precise definition requires subtraction of singularities:
\begin{equation}
e_1(x)=\lim_{\epsilon\to 0}\left( (k+1)\partial_{--}\bar\phi(x)\psi_-(x)\bar\psi_-(x-\epsilon)- i(\partial_{--}\bar\phi(x))^2\phi(x-\epsilon) - \frac{2ki}{\epsilon^{--}}\partial_{--}\bar\phi(x) \right).
\end{equation}
We see that the piece that we subtract is not $\bar{Q}_+$-closed, which already suggests that $e_1(x)$ is probably not in the cohomology. Careful computation of $[\bar{Q}_+, e_1]$, followed by taking the $\epsilon\to0$ limit, shows that:
\begin{equation}
[\bar{Q}_+,e_1]=-(k+1)\left[ (k+1)\phi^k\partial_{--}\psi_- - \frac{1}{2}\psi_{-}\partial_{--}\phi^k \right] -i(k+2)[\bar{Q}_+,\partial_{--}^2\bar\phi].
\end{equation}
So indeed, $e_1$ is not in quantum cohomology. We know that classical observables should be lifted from cohomology in pairs. Therefore, the combination we got on the right, $r_1= (k+1)\phi^k\partial_{--}\psi_- - \frac{1}{2}\psi_{-}\partial_{--}\phi^k$, should be some classical cohomology class which disappears together with $e_1$. And indeed, it is in the classical cohomology, as it is easy to check. Before, we found classical cohomology classes which had a chance of being superconformal primaries, and this $r_1$ was not among them, which suggests that it should be a descendant. Another computation shows that it is indeed a descendant. The lowest component of $\J$ is:
\begin{equation}
j=\J| = \frac{k+1}{2(k+2)}\psi_-\bar\psi_- - \frac{i}{k+2}\phi\partial_{--}\bar\phi,
\end{equation}
it is a $U(1)$ current in the $\N=2$ super-Virasoro. A computation shows that:
\begin{equation}
j_{-1} (\phi^k \psi_-) = :j \phi^k\psi_-:=\frac{i}{k+2}\left((k+1)\phi^k\partial_{--}\psi_- - \psi_-\partial_{--}\phi^k\right) + [\bar{Q}_+,\dots].
\end{equation}
So this new operator, $r_1= (k+1)\phi^k\partial_{--}\psi_- - \frac{1}{2}\psi_{-}\partial_{--}\phi^k$, is actually a superconformal descendant of $\phi^k\psi_-$. One can ask a similar question: what is this $\phi^k\psi_-$? Clearly, it is in the classical cohomology. But in fact, $\phi^k\psi_-=\frac{1}{k+1}[Q_-,\phi^{k+1}]$, and recall that we have a relation $\phi^{k+1}=0$ in the classical cohomology. Therefore $\phi^k\psi_-$ also vanishes in the classical cohomology. So we have discovered the following: classically, we have cohomology classes $e_1$ and $r_1$, but quantum-mechanically, we have $[\bar{Q}_+,e_1]=r_1$. And this $r_1$ is a superconformal descendant of $\phi^k \psi_-$, which is actually zero in the classical cohomology.

This might look confusing -- how is it possible that a superconformal descendant of zero is not zero? The resolution of this apparent paradox is that, actually, super-Virasoro algebra does not act in the classical cohomology. It only acts in the quantum cohomology by the OPE with the stress-energy supercurrent $\J$, while there is no notion of OPE in the classical cohomology. Therefore, there is no contradiction between the facts that $\phi^k\psi_-$ vanishes in the classical cohomology, while its superconformal descendant $r_1$ does not vanish classically. The fact that latter is a descendant of the former is borrowed from the chiral algebra in the quantum cohomology. And in the quantum cohomology, because of this relation, both of them indeed have to vanish. This is quite satisfactory, because it also explains why $r_1$ should be lifted from the classical cohomology -- because it vanishes in quantum chiral algebra!

In fact, by taking all possible superconformal descendants of the relation $\phi^k=0$, we will get a lot of (probably, infinitely many) operators which vanish in the quantum cohomology but represent non-vanishing classical cohomology classes. They all should be lifted from the cohomology through the mechanism which we have just described.

Also, it is not hard to convince oneself that not only $\mathcal{E}^{0,1}$, but all operators $\mathcal{E}^{0,n}$, $n>0$ get lifted from the cohomology at quantum level for the same reasons. Clearly, there is some interesting (or at least non-trivial) mathematical structure in how classical cohomology classes get paired and lifted from the cohomology. It is quite possible that our observables $\mathcal{E}^{s,n}$ and superconformal descendants of $\phi^{k+1}$ are not the only classical cohomology classes involved in this. However, we are not going to study this question here. 

We are only interested in the quantum cohomology here, so the conclusion we need now is that the only primary operators in the cohomology are $1, \vPh, \vPh^2, \dots, \vPh^k$. They, together with the stress-energy supercurrent $\J$, generate the full chiral algebra in the $\bar{Q}_+$-cohomology. One can find that:
\begin{equation}
\J(x_1,\theta_1) \vPh^s(x_2, \theta_2) \sim -\left( \frac{2\theta_{12}^- \bar\theta^-_{12}}{(\mathbf{r}_{12})^2}h_s + \frac{i\theta^-_{12}}{\mathbf{r}_{12}}D_- + \frac{2\theta^-_{12}\bar\theta^-_{12}}{\mathbf{r}_{12}}\partial_{--} + \frac{i}{\mathbf{r}_{12}}q_s \right)\vPh^s,
\end{equation}
where $h_s = \frac{q_s}{2} = \frac{s}{2(k+2)}$. We see that dimensions and charges match exactly our expectations for the $A_{k+1}$ minimal model.

\subsubsection{D and E series of minimal models}
We will not go into much details about the chiral algebras of D and E series of minimal models. Instead we will just look at some of their features, leaving a more detailed study for the future.

The LG models which are expected to flow to $D_{2n+2}$ minimal models in the IR are described by the superpotential:
\begin{equation}
W = XY^2 + \frac{X^{2n+1}}{2n+1}.
\end{equation}

Consider the $n=1$ theory. It has $W = XY^2 + \frac{X^3}{3}$. If we make a change of variables
\begin{align}
V &= \frac{X + Y}{\sqrt{2}},\cr
U &= \frac{X - Y}{\sqrt{2}},
\end{align}
We will get an LG model with $W = \frac{\sqrt{2}}{3}V^3 + \frac{\sqrt{2}}{3}U^3$. This is just a pair of non-interacting $A_2$ models. Thus the theory in the IR is expected to be just $A_2 \otimes A_2$, with the chiral algebra being a tensor product as well. Recall from the previous subsection that, for the $W\propto V^3$, the chiral algebra has only two primaries: the identity $1$ and $V$, and there is also a stress-energy supercurrent $\J_V$. Similarly for the second one: we have $1$ and $U$ as primaries, and we have $\J_U$. By taking the tensor product of these two, we can identify primaries in the chiral algebra of $A_2\otimes A_2$ as:
\begin{equation}
1, V, U, VU, \J_V - \J_U.
\end{equation}
Going back to $X$ amd $Y$, the first four are simply:
\begin{equation}
1, X, Y, X^2 - Y^2.
\end{equation}
Moreover, since in the cohomology $V^2 = U^2 = 0$, these $X$ and $Y$ satisfy relations in the chiral algebra:
\begin{align}
X^2 + Y^2&=0,\cr
XY&=0,
\end{align}
which are just the relations of the chiral ring, so we get the familiar result (explained in Section \ref{SCFT}) that operators from the chiral ring of the $\N=(2,2)$ theory are primaries of the chiral algebra. However, we have an extra primary operator of dimension 1:
\begin{equation}
\P = \J_V - \J_U,
\end{equation}
which is not part of the chiral ring. The existence of this extra primary current in the cohomology was already noticed in \cite{Moh}, where the author also conjectured that every $D_{2n+2}$ model has, in addition to the generators of the chiral ring, a single dimension-$n$ primary in the cohomology.

We are not going to study $n>1$ cases here. The only thing we want to mention is that the spectral sequence approach we used for the $A_{k+1}$ models can be clearly generalized to the $D_{2n+2}$ case. For $n>1$, the operator $\bar{D}_+$ will split as a sum of three terms:
\begin{equation}
\bar{D}_+ = d_0 + d_1 + d_2,
\end{equation}
where $d_0$ corresponds to the zero superpotential, $d_1$ takes into account the effect of $XY^2$ term in the superpotential, and $d_2$ encodes the effect of $X^{2n+1}$ interaction. It should be possible, though more technical than in the $A_{k+1}$ case, to compute the cohomology using this splitting and check the conjecture made in \cite{Moh}.

Finally, a small remark about the $E$ series. The models $E_6$ and $E_8$ correspond to superpotentials $X^3 + Y^4$ and $X^3 + Y^5$. Therefore, their chiral algebras are immediately identified as those of $A_2\otimes A_3$ and $A_2\otimes A_4$ respectively. Therefore, they will also contain extra primary operators, in addition to the chiral ring elements. The $E_7$ model has:
\begin{equation}
W= X^3 + XY^3,
\end{equation}
therefore it has to be studied separately. In this case again we will have:
\begin{equation}
\bar{D}_+ = d_0 + d_1 + d_2,
\end{equation}
where $d_0$ is a $\bar{D}_+$ operator in the theory of two free chiral superfields without any superpotential, $d_1$ takes into account the $X^3$ term and $d_2$ takes care of $XY^3$. It is clear that at the second step of the spectral sequence computation, when we consider the cohomology of $d_1$, we will essentially get the cohomology of the $A_2$ model multiplied by the free theory described by the chiral superfield $Y$. Computing the cohomology of $d_2$ at the next step then becomes much simpler, since we already know the answer for $A_2$.

\section{Discussions and further directions}
We have only scratched the surface of the subject, demonstrating some general properties of chiral algebras of $\N=(0,2)$ theories and giving several simple examples. The most general property was the RG invariance of the answer, which makes chiral algebras interesting objects to study in the context of dualities.

One obvious extension of this work would be to get a better description of chiral algebras of $\N=(2,2)$ LG models with quasi-homogeneous (or even general) polynomial superpotentials. Our treatment allowed us to find answers in some cases, but it would be much nicer to have a more general result, which would associate chiral algebra to any polynomial superpotential. It would also be useful to find some classes of $\N=(0,2)$ models in which the chiral algebra could be described completely.

But the most interesting and immediate extension is, of course, the application of chiral algebras to gauge theories. If the LG model has some flavor symmetry, one can gauge it by coupling to gauge multiplets. One can argue that perturbatively, the way this gauging is implemented in the chiral algebra is as follows. If $G$ is the gauge group, one should first take the $G$-invariant subalgebra of the ungauged chiral algebra, then tensor multiply it by the ``small'' $bc$-system of dimension $(1,0)$ (where ``small'' means that zero mode of $c$ is excluded from the algebra). The ungauged chiral algebra has a current in it which corresponds to the flavor symmetry we want to gauge. Using this current and the $bc$-ghosts, one can construct a BRST operator. The condition of its nilpotency is precisely the condition that there is no gauge anomaly, i.e., that the symmetry we want to gauge really can be gauged. Then we have to compute the cohomology of this BRST operator. The answer is the gauged chiral algebra. (This procedure is identical to the gauging of chiral algebras of 4D $\mathcal{N}=2$ theories from \cite{BR}.)

This procedure seems to hold in perturbation theory. One way to argue it is by writing equations of motion of the gauge theory and, similar to what we did in this paper, computing the cohomology of $\bar{D}_+$ using perturbation theory (or spectral sequence) in gauge coupling. This approach is somewhat ugly, but it allows to argue that the answer is as we claimed above. Another, more conceptual proof would be to define the gauge theory using the BRST formalism and the holomorphic gauge $v_{++}=0$. This would give the action:
\begin{equation}
S = S_0 + \{Q_B,\Psi\}=S_0 + \int \d^2x\, l^A v^A_{++} + \int \d^2x\, b^A\mathcal{D}_{++} c^A,
\end{equation}
where $l^A$ is the auxiliary field implementing gauge $v^A_{++}=0$, and we added Faddeev-Popov ghosts. One can extend supersymmetry to act trivially on ghosts. Then the supercharge $\bar{Q}_+$ and the BRST charge $Q_B$ anticommute: $\{\bar{Q}_+, Q_B\}=0$, and we really have two commuting complexes. The theory is defined as the cohomology of $Q_B$, and within that cohomology we want to find the chiral algebra in the cohomology of $\bar{Q}_+$. Since the complexes commute, we could first find the cohomology of $\bar{Q}_+$, and then compute the cohomology of $Q_B$. It is quite nice to discover that the gauging procedure we explained above arises in this way. However, some technical details still have to be clarified.

A question of utmost importance is to understand how the gauging procedure should be modified to account for non-perturbative effects, such as instantons.

Another extension, which is also important for gauge theories, is to study models without R-symmetry. We can easily find gauge theories with anomalous R-symmetry. In case they are constructed by gauging some LG models that have (right-handed) R-symmetry, it becomes natural to ask what happens to their chiral algebra during gauging that manifests the breaking of R-symmetry.

\acknowledgments

The author would like to thank E. Witten for helpful discussions, as well as B. Le Floch, I. Melnikov and E. Sharpe for useful comments on the draft.

\end{document}